\documentclass[12pt]{article}
\usepackage{epsfig,amssymb}
\usepackage{latexsym}

\newcommand{\bq}{\begin{equation}}
\newcommand{\eq}{\end{equation}}
\newcommand{\bqa}{\begin{eqnarray}}
\newcommand{\eqa}{\end{eqnarray}}
\newcommand{\ben}{\begin{enumerate}}
\newcommand{\een}{\end{enumerate}}
\newcommand{\bc}{\begin{center}}
\newcommand{\ec}{\end{center}}
\newcommand{\bqb}{\begin{eqnarray*}}
\newcommand{\eqb}{\end{eqnarray*}}

\def\pr#1#2#3{ Phys. Rev. ${\bf{#1}}$:#2 (#3)}
\def\prl#1#2#3{ Phys. Rev. Lett. ${\bf{#1}}$:#2 (#3)}
\def\pl#1#2#3{ Phys. Lett. ${\bf{#1}}$:#2 (#3)}
\def\prep#1#2#3{ Phys. Rep. ${\bf{#1}}$:#2 (#3)}
\def\rmp#1#2#3{ Rev. Mod. Phys. ${\bf{#1}}$:#2(#3)}

\def\np#1#2#3{ Nucl. Phys. ${\bf{#1}}$:#2 (#3)}
\def\zp#1#2#3{ Z. f. Phys. ${\bf{#1}}$:#2 (#3)}
\def\jhep#1#2#3{ JHEP ${\bf{#1}}$:#2 (#3)}
\def\epj#1#2#3{ Eur. Phys. J. ${\bf{#1}}$:#2 (#3)}

\def\cpc#1#2#3{Comput. Phys. Commun. ${\bf{#1}}$:#2 (#3)}

\def\aop#1#2#3{Annals of Phys. ${\bf{#1}}$:#2 (#3)}

\def\eg{{\it e.g. ~}}

\def\etal{{\it et.al.~}}

\global\nulldelimiterspace = 0pt

\def\L{ {\cal L }}

\def\swd{s^2_W}

\def\mwd{m_W^2}

\def\tchi{\tilde \chi}

\begin{document}
\pagenumbering{arabic}
\thispagestyle{empty}
\def\thefootnote{\fnsymbol{footnote}}
\setcounter{footnote}{1}

\begin{flushright}
November 2003\\
PM/03-27\\
THES-TP 2003-04 \\
hep-ph/0311076.\\

 \end{flushright}
\vspace{2cm}
\begin{center}
{\Large\bf Neutralino pair production in a $\gamma \gamma $
Collider\footnote{Programme
d'Actions Int\'egr\'ees Franco-Hellenique, Platon 04100 UM}}.
 \vspace{1.5cm}  \\
{\large G.J. Gounaris$^a$, J. Layssac$^b$, P.I. Porfyriadis$^a$
and F.M. Renard$^b$}\\
\vspace{0.2cm}
$^a$Department of Theoretical Physics, Aristotle
University of Thessaloniki,\\
Gr-54124, Thessaloniki, Greece.\\
\vspace{0.2cm}
$^b$Physique
Math\'{e}matique et Th\'{e}orique,
UMR 5825\\
Universit\'{e} Montpellier II,
 F-34095 Montpellier Cedex 5.\\

\vspace*{1.cm}

{\bf Abstract}
\end{center}

We present a complete 1-loop study of the process
$\gamma \gamma \to \tchi^0_i \tchi^0_j$
and the predicted cross section in a $\gamma\gamma$ Linear Collider.
A suitable numerical code PLATONlc, valid for any set of real
MSSM parameters, is released. This study and  code are
complementary to those suitable for Dark Matter
detection  through  the inverse
process $\tchi^0_i \tchi^0_j \to \gamma\gamma$ describing
neutralino-neutralino annihilation at rest, which were presented previously.
If SUSY is realized in Nature, both codes
should be very helpful in future astrophysical and collider
studies of the neutralino sector.

\vspace{0.5cm}

\def\thefootnote{\arabic{footnote}}
\setcounter{footnote}{0}
\clearpage

\section{Introduction}

A tacit candidate particle for the dominant contribution to
  the cold Dark Matter (DM) that apparently constitutes  almost a third
of the energy density of our Universe is, of course,
 the lightest neutralino(s)  predicted in an R-parity conserving
 minimal supersymmetric model (MSSM) \cite{Spergel, Kamio-rep, DMrev}.
Provided the MSSM couplings have the appropriate values,
the  annihilation processes
$\tchi_1^0 \tchi_1^0 \to \gamma \gamma,~ \gamma Z$ could
produce observable rates of very energetic photons
coming from the center of our Galaxy
\cite{nnDM1, nnDM2, nngammaZ, nnDM3, DM-obs}.

Such observable     photon rates could be realized  due to  a large
Wino or Higgsino component for the lightest neutralino $\tchi_1^0$,
and possibly also from resonance effects induced by the $A^0$ or $H^0$
masses \cite{nnDM1, nnDM2, nngammaZ, nnDM3}. In any case,
the observability of such  halo galactic photons
depends  so strongly  on the MSSM parameters,
that even the non-observation of any  signal could produce
useful constraints.

Complementary, and in principle much more detail information
on the neutralino properties could be obtained by studying
$e^-e^+\to \tchi_i^0 \tchi_j^0 $ and
$\gamma \gamma \to \tchi_i^0 \tchi_j^0 $ at a Linear Collider (LC) \cite{LC},
for any neutralino pair $(i,j=1,...,4)$ accessible by the available
energy. Since the  first of these processes  is realized
already at the tree level, studying its signatures  should eventually supply
most of the experimentally accessible  information on
the neutralinos   \cite{eenn, LeMouel}.

Nevertheless, the study
of the 1-loop process $\gamma \gamma \to \tchi_i^0 \tchi_j^0 $
in a $\gamma \gamma$ Linear Collider ($LC_{\gamma \gamma}$) \cite{Laser-LC},
is also useful, since it directly tests in
an Earthy experiment, exactly the same process as in the Dark Matter (DM)
searches. Compared to the DM studies,
the extra advantage of  Collider measurements  though,
is  that they   can be done   for  any neutralino
pair - even  unstable ones, and be performed over a considerable range of
 energies.
Furthermore, depending on the polarizations of the $e^\pm$-beams and
the laser photon beams, six different "cross-section-like" observables
are in principle available, even if we sum over all possible helicities
of the final neutralinos; see \eg \cite{ggV1V2}.
This is much richer than in the case
of DM studies where only the unpolarized total
$\tchi_i^0 \tchi_j^0 \to \gamma \gamma $
cross section for the cosmologically stable $\tchi_i^0$
neutralinos, very close to threshold, is relevant.

The purpose of this  work is to present such a study
based on the complete set of the contributing
1-loop diagrams. In Section 2 we discuss the general
properties of the amplitudes of the processes
$\gamma \gamma \to \tchi_i^0 \tchi_j^0 $ and its reverse,
as well as those of the $LC_{\gamma \gamma}$-observables,
when the neutralino helicities are summed over.
The results for these observables in
an extensive set of MSSM benchmark models are discussed in
Section 3, where the released PLATONlc code is also presented.
Finally Section 4 contains the conclusions.

\section{The processes  $\gamma \gamma \leftrightarrow \tchi_i^0 \tchi_j^0 $}

We denote by $F^{ij}_{\lambda_1,\lambda_2;\mu_1,\mu_2}(\theta)$, the
invariant amplitudes for the process
 $\chi^0_i\chi^0_j~\to~\gamma\gamma$  at a center-of-mass scattering angle
 $\theta$. Here $(\lambda_i, \lambda_j)$ denote the helicities of
  the  neutralinos $(\tchi^0_i, \tchi^0_j)$
  (with $i,j=1,...,4$ being the neutralino counting indices),
 and $(\mu_1, \mu_2)$  the helicities of the
 photons\footnote{Thus  $\lambda_{i,j}=\pm1/2$ and $\mu_{1,2}=\pm 1$,
so that $(-1)^{\mu_1-\mu_2}=1$.}.
The amplitudes for the reverse process $\gamma\gamma~\to~\chi^0_i\chi^0_j$
are written as $\tilde F^{ij}_{\mu_1,\mu_2;\lambda_1,\lambda_2}(\theta)$.

Following  the Jacob-Wick conventions \cite{JacobW},
the $\chi^0_i\chi^0_j$-fermion  antisymmetry
implies\footnote{These relations do not agree with the
antisymmetry condition used in \cite{China1}.}
\bqa
&& F^{ij}_{\lambda_1,\lambda_2;\mu_1,\mu_2}(\theta)=(-1)^{\mu_1-\mu_2}~
F^{ji}_{\lambda_2,\lambda_1;\mu_1,\mu_2}(\pi-\theta) ~~,
\nonumber \\
&& \tilde F^{ij}_{\mu_1,\mu_2;\lambda_1,\lambda_2}(\theta)=(-1)^{\mu_1-\mu_2}~
\tilde F^{ji}_{\mu_1,\mu_2;\lambda_2,\lambda_1}(\pi-\theta) ~~,
\label{fermion-antisymmetry}
\eqa
 while the $\gamma \gamma$-boson symmetry requires
\bqa
&& F^{ij}_{\lambda_1,\lambda_2;\mu_1,\mu_2}(\theta)=(-1)^{\lambda_1-\lambda_2}~
F^{ij}_{\lambda_1,\lambda_2;\mu_2,\mu_1}(\pi-\theta) ~~,
\nonumber \\
&& \tilde F^{ij}_{\mu_1,\mu_2;\lambda_1,\lambda_2}(\theta)
=(-1)^{\lambda_1-\lambda_2}~
\tilde F^{ij}_{\mu_2,\mu_1;\lambda_1,\lambda_2}(\pi-\theta)~~.
\label{boson-symmetry}
\eqa

If the MSSM breaking parameters and  the Higgs parameter
$\mu$ are real,
then time reversal and CP invariance hold for the neutralino processes,
implying also
\bqa
&& \tilde F^{ij}_{\mu_1,\mu_2;\lambda_1,\lambda_2}(\theta)=
F^{ij}_{\lambda_1,\lambda_2;\mu_1,\mu_2}(\theta) ~~ .
\label{time-invariance} \\
&& F^{ij}_{-\lambda_1,-\lambda_2;-\mu_1,-\mu_2}(\theta)=
(-1)^{\lambda_1-\lambda_2-(\mu_1-\mu_2)}~\eta_i\eta_j
F^{ij}_{\lambda_1,\lambda_2;\mu_1,\mu_2}(\theta) ~~,
\nonumber \\
&& \tilde F^{ij}_{-\mu_1,-\mu_2;-\lambda_1,-\lambda_2}(\theta)=
(-1)^{\lambda_1-\lambda_2-(\mu_1-\mu_2)}~\eta_i\eta_j
\tilde F^{ij}_{\mu_1,\mu_2;\lambda_1,\lambda_2}(\theta) ~~,
\label{CP-invariance}
\eqa
where $\eta_j=\pm 1$ is the CP-eigenvalue of the
neutralino\footnote{We follow the same notation as in \eg \cite{LeMouel}.}
$\tchi^0_j$.

Combining  (\ref{fermion-antisymmetry},
\ref{boson-symmetry}, \ref{CP-invariance}), we get
\bqa
&& F^{ij}_{\lambda_1\lambda_2;\mu_1\mu_2}(\theta)=
(-1)^{\mu_1-\mu_2 +\lambda_2-\lambda_1}
F^{ji}_{\lambda_2\lambda_1;\mu_2\mu_1}(\theta)=
\eta_i \eta_j F^{ji}_{-\lambda_2,-\lambda_1;-\mu_2, -\mu_1}(\theta) ~~,
\nonumber  \\
&& \tilde F^{ij}_{\mu_1\mu_2; \lambda_1\lambda_2}(\theta)=
(-1)^{\mu_1-\mu_2 +\lambda_2-\lambda_1}
\tilde F^{ji}_{\mu_2\mu_1; \lambda_2\lambda_1}(\theta)=
\eta_i \eta_j \tilde F^{ji}_{-\mu_2, -\mu_1;-\lambda_2,-\lambda_1}(\theta)~~,
\label{neutralino-photon-symmetry}
\eqa
where the first part comes from (\ref{fermion-antisymmetry},
\ref{boson-symmetry}) alone, while for the last part the CP-invariance
relation (\ref{CP-invariance}) is also used. On the basis of
(\ref{neutralino-photon-symmetry}) we could select
$F_{++++}$, $F_{++--}$, $F_{+++-}$, $ F_{+-++}$, $F_{+-+-}$, $F_{+--+}$
as a possible set of independent helicity amplitudes at
each specific angle.

\begin{figure}[p]
\vspace*{-3cm}
\[
\hspace{-1.cm}\epsfig{file=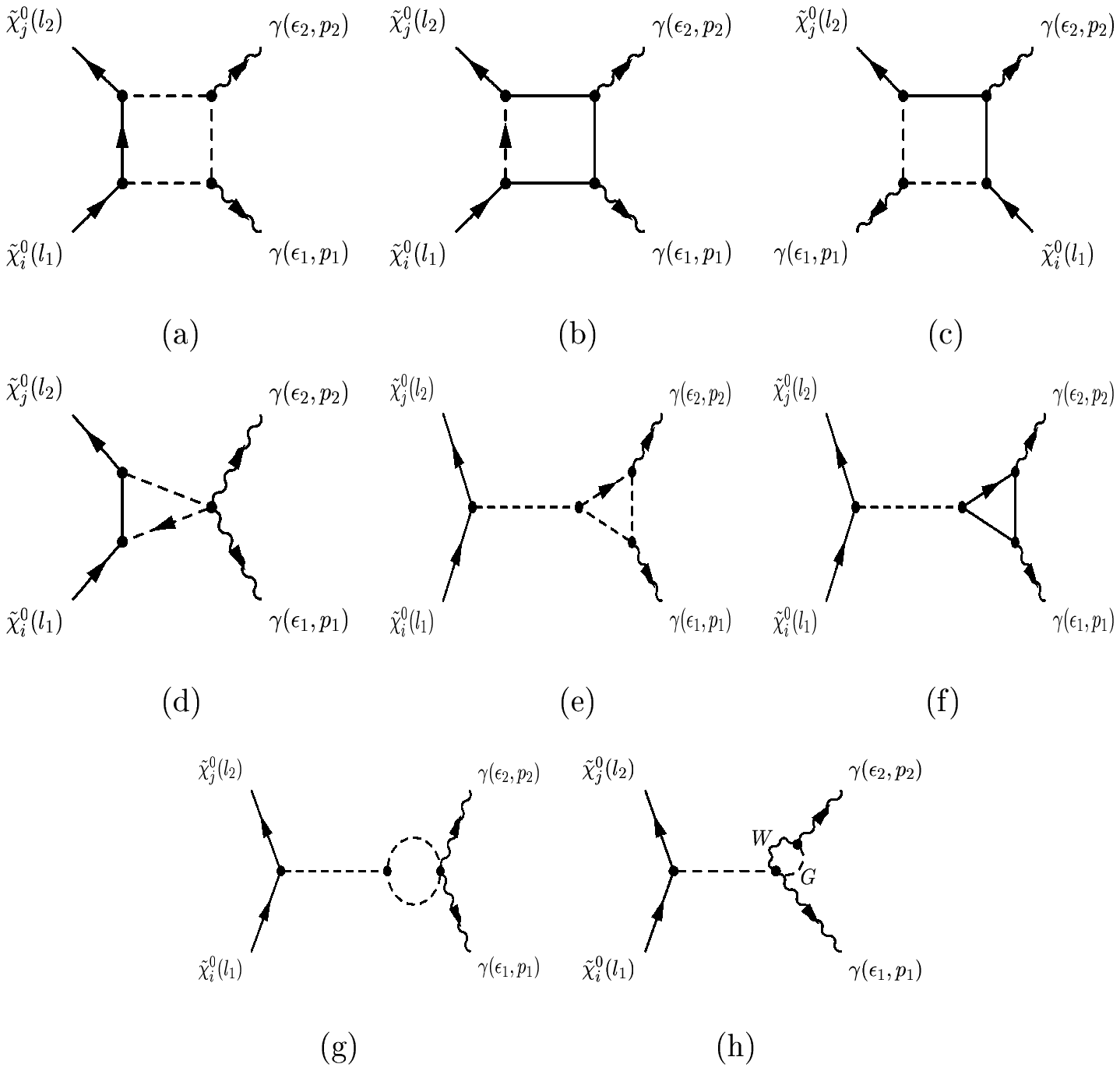,height=13.cm, width=13.cm}
\]
\caption[1]{ Feynman diagrams for $\chi^0_i\chi^0_j \to \gamma\gamma$.
Full internal lines denote fermionic exchanges; while
broken internal lines  denote either scalar or gauge exchanges,
 except in the diagram (h), where the
 W and Goldstone exchanges are  indicated explicitly.}
\label{Diagrams}
\end{figure}

Restricting to real MSSM parameters from here on,
it is thus  sufficient to
only consider the amplitudes of the
 process $\tchi^0_i\tchi^0_j \to \gamma\gamma $, for both DM and
 $LC_{\gamma\gamma}$ studies,
 using  the  one-loop diagrams generically described
in Fig.\ref{Diagrams}, in the 't-Hooft-Feynman gauge.
The  types of  particles running
clockwise inside each  loop in Fig.\ref{Diagrams}  are:
box\footnote{In naming the particle-strings in the box-loop,
we always start (moving clockwise) from the line ending to the
$\tchi^0_j$ vertex.}
(a):  $(fSSS)$, $(fWWW)$, $(fSSW)$, $(fSWW)$, $(fSWS)$,
$(fWWS)$; box (b):   $(Sfff)$, $(Wfff)$;
box (c):  $(SffS)$,  $(WffW)$, $(SffW)$, $(WffS)$;
initial triangle\footnote{In naming these particle-strings
in Fig.\ref{Diagrams}d,  we always start from the line ending to the
$\tchi^0_i$-vertex.} (d):  $(SfS)$, $(WfW)$;
final triangle\footnote{The particle-string names
in Fig.\ref{Diagrams}e,f,   always start from the line leaving  the
 vertex where the s-channel exchanged neutral particle
 $Z$,   $h^0$, $H^0$, $A^0$ or  $G^0$ ends.}
 (e):   $(WWW)$, $(SSS)$, $(SWW)$, $(SSW)$;
final triangle  (f): $(fff)$; and bubbles (g): $(SS)$, $(WW)$; (h): $(WS)$.
By $S$ we denote the  scalar exchanges
(Higgs, Goldstone and  sfermions);
and by $f$  the  fermionic ones  (leptons, quarks and inos).
Bubbles (g) and final
triangles (e) and (f), are connected to the initial $\tchi^0_i \tchi^0_j$ state
through an intermediate $Z$,  or neutral  Higgs or
Goldstone boson $h^0$, $H^0$, $A^0$,  $G^0$.

For  book-keeping  the Majorana nature of the neutralinos,
we always describe $\tchi^0_i$ by a positive energy Dirac wavefunction,
 and $\tchi^0_j$ by a negative energy one. We then note that the
 triangle and bubble diagrams in Fig.\ref{Diagrams},
 separately satisfy  the  neutralino and photon symmetry   conditions
 (\ref{fermion-antisymmetry}, \ref{boson-symmetry}).
 For the  separate box-diagrams  in Fig.\ref{Diagrams} though,
 the neutralino  exchange symmetry (\ref{fermion-antisymmetry})
 is only satisfied after  adding
   to each  box of Fig.\ref{Diagrams}, the corresponding photon exchanged one.
   The calculation of these later  boxes  may be
     avoided though, by imposing (\ref{boson-symmetry}) to the contribution of
   each of the boxes in Fig.\ref{Diagrams}.
  Requiring then the  validity  of  (\ref{fermion-antisymmetry})
   for each  such box contribution,
 provides a stringent test of the computation.
 We have checked that all these constraints are exactly satisfied.

We have derived explicit analytic expressions for the contributions of
the various  diagrams of Fig.\ref{Diagrams} in terms of
Passarino-Veltman functions \cite{Passarino}.
These  expressions are so lengthy though,
that it is useless  to present them
explicitly. We have therefore instead chosen to release a
numerical code, which calculates the differential cross sections
defined below,  for various photon polarizations
in any MSSM model with real  parameters at the
electroweak scale; see discussion below.

Before turning to this  though, we add two remarks concerning the
behavior of the amplitudes  near threshold,
and  at very high energies and angles, respectively.
Very close to threshold, the relative orbital angular momentum  $l$ of the
neutralino pair  $\tchi_i^0 \tchi_j^0$ is, of course, vanishing.
Because of Fermi statistics,
if $i=j$, the neutralino-pair can then only exist
in an $^1S_0$-state corresponding to a total neutralino spin $S_{\rm tot}=0$;
 while if $i\neq j$, $S_{\rm tot}=1$ is also possible. In both cases,
 the CP eigenvalues of the $\tchi_i^0 \tchi_j^0$-pair is
\[
~~~CP= -(-1)^l\eta_i\eta_j ~~~.
\]
Thus, for $l=0$, $S_{\rm tot}=0$ and $\eta_i\eta_j=1$
($\eta_i\eta_j=-1$),  the neutralino pair
has the same quantum numbers as the $A^0$ ($H^0$) neutral Higgs-state,
leading to a resonance enhancement for the $\lambda_i=\lambda_j $-amplitude,
 provided   the appropriate neutral Higgs mass is close to the
 sum of the neutralino  masses\footnote{Since the neutralino
 masses are  heavier than  the $Z$-mass in all  contemplated MSSM models,
    there can never be an analogous enhancement due to the
intermediate $Z$ boson in the $i\neq j$,
    $S_{\rm tot}=1$ case.}. This is induced by the
    diagrams in Figs.\ref{Diagrams}e,f,g,h.

Next, concerning the behavior at very high energy
($s \sim |t| \sim |u| \gg M_{SUSY}$-scale\footnote{For such
kinematical situations:~
 $t\sim -s(1-\cos\theta)/2$ and $u \sim -s(1+\cos\theta)/2$.}),
 the dominant leading-log
contribution to the neutralino production  amplitude is found to be
generated by  the diagrams in Figs.\ref{Diagrams}a,b,c,d when   the internal
loops involve  chargino and $W$-exchanges. This is given by
\bq
F^{ij}_{\lambda_1\lambda_2; \mu_1\mu_2}=
\tilde F^{ij}_{\mu_1\mu_2;\lambda_1\lambda_2}~\simeq ~
- \frac{\alpha^4 }{2\swd }\sin\theta
\left [ C_R  {1+\eta_i\eta_j\over2}\L_R +C_I  {1-\eta_i\eta_j\over2}
\L_I \right ]+\cdot \cdot \cdot ~~, \label{leading-F}
\eq
where the dots stand for subleading  constant and $1/s$-terms, while
\bqa
\L_R & = &
\Big [(\mu_2-\mu_1)(\lambda_1-\lambda_2)^2+2\mu_1\mu_2
(\lambda_1-\lambda_2)\cos\theta \Big ]
\Big [\frac{s}{u} \Big (\ln^2 \frac{|t|}{\mwd}-\ln^2 \frac{s}{\mwd}
+2i\pi\ln \frac{s}{\mwd} \Big)
\nonumber \\
&+ & \frac{s}{t} \Big (\ln^2 \frac{|u|}{\mwd}-\ln^2 \frac{s}{\mwd}
+2i\pi\ln \frac{s}{\mwd}\Big ) \Big  ]
+  (1+\mu_1\mu_2)(\lambda_1-\lambda_2) \Big [
\frac{s}{u}\Big (\ln^2 \frac{|t|}{\mwd} -\ln^2 \frac{s}{\mwd}
\nonumber \\
&+ &2i\pi\ln \frac{s}{\mwd} \Big )
+\frac{s}{t}\Big (\ln^2 \frac{s}{\mwd}
-2i\pi\ln \frac{s}{\mwd}-\ln^2 \frac{|u|}{\mwd} \Big )
+2 \left (\ln^2 \frac{|t|}{\mwd}-\ln^2 \frac{|u|}{\mwd}\right ) \Big ] ~,
\label{leading-LR} \\
\L_I & = &
\Big [(\mu_2-\mu_1)(\lambda_1-\lambda_2)+2\mu_1\mu_2
(\lambda_1-\lambda_2)^2\cos\theta \Big ]
\Big [\frac{s}{u}  \Big (\ln^2 \frac{|t|}{\mwd}-\ln^2 \frac{s}{\mwd}
+2i\pi\ln \frac{s}{\mwd}\Big )
\nonumber \\
&+ & \frac{s}{t} \Big (\ln^2 \frac{|u|}{\mwd}-\ln^2 \frac{s}{\mwd}
+2i\pi\ln \frac{s}{\mwd}\Big ) \Big  ]
+  (1+\mu_1\mu_2)(\lambda_1-\lambda_2)^2 \Big [
\frac{s}{u}\Big (\ln^2 \frac{|t|}{\mwd} -\ln^2 \frac{s}{\mwd}
\nonumber \\
&+ &2i\pi\ln \frac{s}{\mwd}\Big )
+ \frac{s}{t}\Big (\ln^2 \frac{s}{\mwd}
-2i\pi\ln \frac{s}{\mwd}-\ln^2 \frac{|u|}{\mwd} \Big )
+2 \left (\ln^2 \frac{|t|}{\mwd}-\ln^2 \frac{|u|}{\mwd}\right ) \Big ] ~,
\label{leading-LI}
\eqa
and
\bqa
C_R & = & 2Re(Z^N_{2j}Z^{N*}_{2i})+{1\over2}Re(Z^N_{4j}Z^{N*}_{4i}
+Z^N_{3j}Z^{N*}_{3i}) ~,\nonumber \\
C_I &= & 2iIm(Z^N_{2j}Z^{N*}_{2i})+{i\over2}Im(Z^N_{4j}Z^{N*}_{4i}
+Z^N_{3j}Z^{N*}_{3i})~,
\label{CRCI}
\eqa
with   $Z^N_{\alpha j}$ describing the neutralino mixing matrix in the
notation of \cite{Rosiek}. As expected from \cite{Verzegnassi},
gauge invariance eliminates all isolated
single-log and log-squared terms
in (\ref{leading-F}, \ref{leading-LR}, \ref{leading-LI}), so that the
only allowed leading-log contribution consists in differences
of log-squared terms, i.e. a single $\ln (s/ m^2_W)$ multiplied by
angular dependent but energy independent logarithmic
coefficients  for the real part of the amplitudes, and
 simple  $\pi \ln (s/ m^2_W)$-terms for the imaginary parts.
At large angles, this has the tendency to enhance
the imaginary parts of the amplitudes, as compared to the real parts.
Such an asymptotic   behavior has also been observed
(occasionally very strikingly), for the helicity amplitudes of process like
 $\gamma \gamma \to \gamma \gamma, ~Z\gamma, ~ZZ,~ A^0A^0$, always
 being generated  by W-loops  \cite{ggV1V2, ggA0A0}. \par

Note (from the above expressions of $\L_R$ and $\L_R$)
that, in the high energy limit (when masses are neglected and
$s+t+u=0$) the only non vanishing helicity amplitudes
are those with $\lambda_1=-\lambda_2$ and $\mu_1=-\mu_2$, i.e.
$F_{+-+-}$ and $F_{+--+}$ and the ones related by symmetry relations
eq.(\ref{fermion-antisymmetry}-\ref{neutralino-photon-symmetry}).
This is due to the dominance of box diagrams with chirality
conserving gauge couplings of $W$ bosons.\par

In practice these asymptotic expressions have some interest only
when the energy is sufficiently higher than the SUSY scale;
for example in the TeV range if  the SUSY masses are of the order of
a few hundred  GeV. At lower energies, two types of effects
modify the asymptotic behavior. One is purely kinematic and
due to the additional  constant and energy decreasing
($M^2/s$) terms denoted by the dots in (\ref{leading-F}). The other one
is the appearance of chirality violating amplitudes due to box and
triangle diagrams involving scalar couplings
(neutralino-sfermion-fermion and neutralino-neutralino-Higgs
couplings). These amplitudes vanish like $M^2/s$ at high
energies, but at  energies not too far from the SUSY scale,
their relative importance, compared   to the one of the
leading chirality conserving ones, is model-dependent.
In particular, it depends on the neutralino
contents (Bino, Wino and Higgsino mixture)
which controls the relative magnitude of its gauge
and scalar couplings.\par

All these properties can be checked by making a numerical comparison
with the complete results. We come back to the discussion
of these in the next section, where we discuss our
results  for various benchmark models.

\vspace{0.5cm}
We next turn  to the $\gamma\gamma\to \chi^0_i\chi^0_j$ collisions
in a $\gamma \gamma $ collider  $LC_{\gamma \gamma}$ \cite{Laser-LC}
realized  through backscattering of laser photons in an $e^-e^+$
 LC \cite{LC}. For  real MSSM parameters,
 where CP-invariance holds, the general form of the
neutralino production cross section is \cite{ggV1V2}:
\bqa
{d\sigma(\gamma \gamma \to \tchi^0_i \tchi^0_j)
\over d\tau d\cos\theta}&=&{d \bar L_{\gamma\gamma}\over
d\tau} \Bigg \{
{d{\sigma}_0\over d\cos\theta}
+\langle \xi_2 \xi^\prime_2 \rangle {d{\sigma}_{22}\over d\cos\theta}
\nonumber\\
&&
+\langle\xi_3\rangle  \,{d{\sigma}_{3}\over d\cos\theta}\cos2\phi
+\langle\xi_3^ \prime\rangle {d\sigma_3^\prime\over d\cos\theta}\cos2\phi^\prime
\nonumber\\
&&
+\langle\xi_3 \xi_3^\prime\rangle
\Big [~{d{\sigma}_{33}\over d\cos\theta}
\cos (2[\phi+\phi^\prime])
+{d{\sigma}^\prime_{33}\over
d\cos\theta^*}\cos (2[\phi- \phi^\prime])\Big ]
\nonumber\\
&&+ \langle\xi_3 \xi_2^\prime\rangle
{d{\sigma}_{23}\over d\cos\theta} \sin 2 \phi
+ \langle\xi_2 \xi^\prime_3\rangle
{d{\sigma^\prime}_{23}\over d\cos\theta}\sin 2\phi^\prime
\Bigg \} \ \ .
\label{sigpol}
\eqa
 In (\ref{sigpol}),  $\tau=s /s_{ee}$, with $s_{ee}$ being the square of
the $e^-e^+$  center-of-mass energy and
$s\equiv s_{\gamma \gamma}=s_{\tchi^0_i \tchi^0_j}$
denoting    the corresponding quantity for the
produced neutralino pair.
The quantity  $ d \bar L_{\gamma\gamma}/d\tau$ describes the photon-photon
luminosity per unit $e^-e^+$ flux, while the parameter-pairs
$(\xi_2, \xi^\prime_2)$, $(\xi_3,~ \xi^\prime_3)$
and $(\phi,~ \phi^\prime)$ describe respectively
 the average helicities, transverse
polarizations and azimuthal angles of the two
 backscattered photons. These are in turn determined
 by the corresponding quantities of the laser photons, and the
 $e^\pm$-polarizations \cite{Laser-LC, ggV1V2}.

Finally, the $\sigma_n$-quantities in (\ref{sigpol})
are defined as
\bqa
 {d \sigma_0\over d\cos\theta}&& =
\left ({C_{ij}\over128\pi s}\right )
\sum_{\lambda_1\lambda_2} [~|\tilde F_{++\lambda_1\lambda_2}|^2
+|\tilde F_{--\lambda_1\lambda_2}|^2+|\tilde F_{+-\lambda_1\lambda_2}|^2
+|\tilde F_{-+\lambda_1\lambda_2}|^2~] ~ , \nonumber\\
&&=\left ({C_{ij}\over64\pi s}\right )
[~|\tilde F_{++++}|^2+ |\tilde F_{++--}|^2
+ |\tilde F_{+++-}|^2+ |\tilde F_{++-+}|^2\nonumber\\
&&
+ |\tilde F_{+-++}|^2 + |\tilde F_{+---}|^2
+ |\tilde F_{+-+-}|^2+ |\tilde F_{+--+}|^2~]~~, \label{sig0} \\[0.5cm]
 {d{\sigma}_{22}\over d\cos\theta}&& =
\left ({C_{ij}\over128\pi s}\right )
\sum_{\lambda_1\lambda_2} [~|\tilde F_{++\lambda_1\lambda_2}|^2
+|\tilde F_{--\lambda_1\lambda_2}|^2-|\tilde F_{+-\lambda_1\lambda_2}|^2
-|\tilde F_{-+\lambda_1\lambda_2}|^2~]   \nonumber\\
&&=\left ({C_{ij}\over64\pi s}\right )
[~|\tilde F_{++++}|^2+ |\tilde F_{++--}|^2
+ |\tilde F_{+++-}|^2+ |\tilde F_{++-+}|^2\nonumber\\
&&
-( |\tilde F_{+-++}|^2 + |\tilde F_{+---}|^2
+ |\tilde F_{+-+-}|^2+ |\tilde F_{+--+}|^2)~]   \label{sig22}~~, \\[0.5cm]
{d{\sigma}_{3} \over d\cos\theta} && =
\left ({-C_{ij}\over64\pi s}\right )  \sum_{\lambda_1\lambda_2}
 [\tilde F_{++\lambda_1\lambda_2}\tilde F^*_{-+\lambda_1\lambda_2}
+\tilde F_{+-\lambda_1\lambda_2} \tilde F^*_{--\lambda_1\lambda_2} ]
 \nonumber \\
&& =\left ({-C_{ij}\over32\pi s}\right )
Re[\tilde F_{++++}\tilde F^*_{+---}
 + \tilde F_{++--}\tilde F^*_{+-++}
 \nonumber \\
&& - \tilde F_{+++-}\tilde F^*_{+--+}-
\tilde F_{++-+}\tilde F^*_{+-+-}]\eta_i\eta_j
~, \label{sig3}  \\[0.5cm]
 {d{\sigma^\prime}_{3}\over d\cos\theta} && =
\left ({-C_{ij}\over64\pi s}\right ) \sum_{\lambda_1\lambda_2}
[\tilde F_{++\lambda_1\lambda_2}\tilde F^*_{+-\lambda_1\lambda_2}
+\tilde F_{-+\lambda_1\lambda_2} \tilde F^*_{--\lambda_1\lambda_2}]
\nonumber\\
&& =\left ({-C_{ij}\over32\pi s}\right )
Re[\tilde F_{++++}\tilde F^*_{+-++}+ \tilde F_{++--}\tilde F^*_{+---}
\nonumber\\
&&  + \tilde F_{+++-}\tilde F^*_{+-+-}+ \tilde F_{++-+}\tilde F^*_{+--+}~]
~, \label{sig3prime} \\[0.5cm]
{d \sigma_{33} \over d\cos\theta}& & =
\left ({C_{ij}\over 32\pi s}\right )
Re[\tilde F_{+---}\tilde F^*_{-+--}+\tilde F_{+-+-}\tilde F^*_{-++-}~]
\nonumber\\
&&=\left ({C_{ij}\over32\pi s}\right )
Re[\tilde F_{+---}\tilde F^*_{+-++}-\tilde F_{+-+-}\tilde F^*_{+--+}]
\eta_i\eta_j  \ , \label{sig33} \\[0.5cm]
{d{\sigma^\prime}_{33}\over d\cos\theta} &&=
\left ({C_{ij}\over 32\pi s}\right )
Re[\tilde F_{++++}\tilde F^*_{--++}+\tilde F_{+++-}\tilde F^*_{--+-}~]
\nonumber\\
&&=\left ({C_{ij}\over32\pi s}\right )
Re[\tilde F_{++++}\tilde F^*_{++--}-\tilde F_{+++-}\tilde F^*_{++-+}]
\eta_i\eta_j ~, \label{sig33prime} \\[0.5cm]
 {d{\sigma}_{23}\over d\cos\theta}& &=
  \left ({-C_{ij}\over 32\pi s}\right ) Im [
\tilde F_{++++}\tilde F^*_{+---}+\tilde F_{++--}\tilde F^*_{+-++}
\nonumber\\
&& -\tilde F_{+++-}\tilde F^*_{+--+}-\tilde F_{++-+}\tilde F^*_{+-+-} ]
\eta_i\eta_j
\label{sig23} ~, \\[0.5cm]
{d{\sigma}_{23}^\prime \over d\cos\theta}&&
=  \left ({C_{ij}\over 32\pi s}\right ) Im [
\tilde F_{++++}\tilde F^*_{+-++}+\tilde F_{++--}\tilde F^*_{+---}
\nonumber\\
&& +\tilde F_{+++-}\tilde F^*_{+-+-}+\tilde F_{++-+}\tilde F^*_{+--+} ]
~,   \label{sig23prime}
\eqa
where the reduction due to the   identity of the final neutralinos
whenever $i=j$, is taken into account through the coefficient
\bqa
 C_{ij}&= &\beta_{ij}\left (1-{\delta_{ij}\over2}\right )~~~, \nonumber  \\
\beta_{ij}&= &\sqrt{\Big [1-{(m_i-m_j)^2\over s}\Big]
\Big[1-{(m_i+m_j)^2\over s}\Big ]} ~~~,
\eqa
and  the c.m. scattering angle $\theta$  varies in the range
$-1 \leq cos\theta \leq  +1 $.

 On the basis of (\ref{fermion-antisymmetry}, \ref{boson-symmetry}) we
 obtain that ${d \sigma_0/ d\cos\theta}$, ${d \sigma_{22}/ d\cos\theta}$,
${d \sigma_{33}/ d\cos\theta}$ and
${{d \sigma^\prime}_{33}/ d\cos\theta}$ are symmetric
 under the interchange
\[
\theta   \leftrightarrow \pi-\theta ~~~,
\]
whereas
\bqa
\frac {d \sigma_3}{ d\cos\theta}\Big |_\theta & =  &
\frac {d \sigma^\prime_3}{ d\cos\theta}\Big |_{\pi-\theta} ~~ ,\nonumber \\
\frac{ d \sigma_{23}}{ d\cos\theta }\Big |_\theta   & = &
 -~ \frac{d \sigma^\prime_{23}}{d\cos\theta}\Big |_{\pi-\theta} ~~ . \nonumber
\eqa

\section{Results}

On the basis of the diagrams in Fig.\ref{Diagrams}, we
have constructed  a numerical code called PLATONlc which calculates the
differential cross sections in (\ref{sig0}-\ref{sig23prime})
for any c.m. energy and any set of real MSSM parameters at the electroweak scale,
using \cite{Oldenborgh}. This code, together with an explanatory
Readme file,  can be downloaded from \cite{plato}.

We have made a run for the 31 benchmark models already used
in the previous DM study  in \cite{nnDM3}. For each model
we have computed both the angular distribution and the
energy dependence of the various cross sections presented in Section 2,
for $\tchi^0_1\tchi^0_1$,
$\tchi^0_1\tchi^0_2$ and $\tchi^0_2\tchi^0_2$ production.

The typical feature of the angular distribution is an isotropy
at low energy, and  a tendency to present some forward-backward
peaking at high energies (see the $1/t$ and $1/u$
terms in eq.(\ref{leading-LR}, \ref{leading-LI})). For  the integrated
cross sections, one finds energy dependent structures
at  $s \sim (m_1+m_2)^2$, with $m_{1,2}$
being the masses of  intermediate  particles
contributing to  the diagrams
of Fig.\ref{Diagrams}, as well as resonance effects due to
diagrams involving $A^0$ or  $H^0$
intermediate neutral Higgs-states.

These features are common to several benchmark models, so we will not
illustrate all of them. Moreover, since the expected luminosities
at the future colliders  will probably be of the the order
of $10^{2}~ \rm{fb}^{-1}$ per year,
we  ignore  those model-cases considered in \cite{nnDM3},
where   the cross sections are much  smaller
 than $\sim 0.1~\rm fb$. We thus present
  results  for the   universal  mSUGRA   benchmark models \cite{mSUGRA-mod}
  $SPS1a_1$,
$SPS4$ \cite{Snowmass},  $AD(fg5)_1$ \cite{Arnowitt}; the  non-universal
mSUGRA  models $CDG_{75}$ and $CDG_{OII}$ \cite{CDG};  and the
 GMSB  models \cite{GMSB-mod}   $SPS7_1$, $SPS8_1$ \cite{Snowmass}.
The grand scale  defining    parameters for all these cases    are shown  in
Table 1.
\begin{table}[hbt]
\begin{center}
{ Table 1: Input  parameters for the used MSSM benchmark models at
the grand scale. Dimensions in GeV. In all cases   $\mu>0$. By
convention, $M_2>0$ is used.}\\
  \vspace*{0.3cm}
\begin{small}
\begin{tabular}{||c|c|c|c|c|c||c|c|c||}
\hline \hline
 & \multicolumn{3}{|c|}{universal mSUGRA } &
\multicolumn{2}{|c||}{non-universal mSUGRA }&
\multicolumn{3}{|c||}{GMSB }\\ \hline
 & $SPS1a_1$  & $AD(fg5)_1$ & $SPS4$ & $CDG_{75}$
 & $CDG_{OII}$&  & $SPS7_1$  & $SPS8_1$     \\ \hline
 $M_1$ &250  & 400  & 300  &-400 & 424  & $M_{mess}$ & 80000 & 120000 \\
$M_2$ & 250 & 400 & 300  &240 & 200   & $M_{SUSY}$ &40000 & 60000 \\
$M_3$ & 250 & 400 &300  & 80 & 40   & $\tan\beta$ &15 & 15 \\
$m_0$ & 100 &220  &400  & 1400 & 1400  & & &  \\
$m_{H_u}$ & 100  &  220& 400 & 1400 & 1400 & & &  \\
$A_0$ &-100  & 0 & 0  & 1000  & 1000   &  & & \\
$\tan\beta$ & 5  & 40  & 50 &  50 & 50  &  & & \\
  \hline \hline
\end{tabular}
 \end{small}
\end{center}
\end{table}

Starting from  the grand-scale parameters in Table 1,
their electroweak scale values  and the $H^0$ and
$A^0$ widths are calculated using the
codes SuSpect \cite{suspect} and HDECAY \cite{HDECAY}. In turn, these
values constitute the input needed for the PLATONlc code \cite{plato}.
The  obtained results are indicated in Figs.\ref{SPS1a-1-fig}-\ref{SPS8-1-fig}
below. In all cases, the sub-figures labelled (a), (b) and (c) describe the
integrated cross sections of Eqs.(\ref{sig0}-\ref{sig23prime}),
in the range $(1^\circ < \theta < 179^\circ )$ for $\tchi^0_1\tchi^0_1$,
$\tchi^0_1\tchi^0_2$ and $\tchi^0_2\tchi^0_2$ production respectively.
Of course, only $\tchi^0_1\tchi^0_2$ and $\tchi^0_2\tchi^0_2$
production are in principle observable, since the lightest
supersymmetric particle LSP $\tchi^0_1$ should be invisible
in an R-conserving theory.
In addition,   sub-figures (d) and (e) give the unpolarized differential
cross section $d\sigma_0/d\cos\theta$ at c.m energies 0.4 and 2 TeV
respectively, for the $\tchi^0_i\tchi^0_j$ production channels
which are energetically accessible.
On the basis of Figs.\ref{SPS1a-1-fig}-\ref{SPS8-1-fig}, we
remark the following:

In  the universal mSUGRA cases of Figs.\ref{SPS1a-1-fig}-\ref{SPS4-fig}
and the non-universal mSUGRA $CDG_{75}$ of Fig.\ref{CDG-75-fig}, only
the $\sigma_0$, $\sigma_{22}$ and occasionally the $\sigma_3=\sigma_3^\prime$
cross sections for $\gamma \gamma \to \tchi^0_2\tchi^0_2 $,
 are sufficiently large to be (in principle)
 observable in a conceivable future Collider. This is because
 $\tchi^0_2\sim \tilde W^{(3)}$ and  $\tchi^0_1\sim \tilde B$,  in
 these  models.
 The largest  and easier to measure cross section is, of course,
 the unpolarized  $\sigma_0$, defined in (\ref{sig0}).
As seen from Figs.\ref{SPS1a-1-fig}-\ref{CDG-75-fig}(c) though,
in an $LC_{\gamma \gamma}$ collider at 0.4 TeV, the $\tchi^0_2\tchi^0_2$-pair
production can only occur in the case of
the $SPS1a_1$ and $CDG_{75}$ models. In all other cases,
$\tchi^0_2$ is too heavy to be produced at 0.4 TeV,
but it can be generated  in a 2TeV $\gamma \gamma$ version of   CLIC;
compare Figs.\ref{SPS1a-1-fig}-\ref{CDG-75-fig}(d).
As seen there, $d\sigma_0/d\cos\theta$ is  very flat at 0.4TeV,
and develops a moderate forward-backward peak already at 2TeV.

Fig.\ref{CDG-OII-fig} concerns the $OII$ non-universal mSUGRA
model of \cite{CDG}, where  $\tchi^0_1$ is predominantly a Wino
with some appreciable Higgsino components, while $\tchi^0_2$ is
predominantly a Bino. Depending on the $LC_{\gamma \gamma}$ energy and
flux, both $\tchi^0_1\tchi^0_2$ and
$\tchi^0_2\tchi^0_2$ productions may be observable. A special feature
of this model is that  $\tchi^0_1\tchi^0_2$
production is  predicted to be generally more copious than the
$\tchi^0_2\tchi^0_2$ one.  Although,
$d\sigma_0/d\cos\theta$ is  very flat at 0.4TeV for all channels,
it develops  strong   forward-backward peaks by the time the
energy reaches  2TeV; compare subfigures (c) and (d).

Concerning  the GMSB models \cite{GMSB-mod}
in Figs.\ref{SPS7-1-fig},\ref{SPS8-1-fig}, we remark that their
general structure is reminiscent of the results in the mSUGRA
models of Figs.\ref{SPS1a-1-fig}-\ref{SPS4-fig}.
Finally, no results are presented for the AMSB models \cite{AMSB-mod, Snowmass},
since the almost exact relations $\tchi^0_1\sim \tilde W$
and $\tchi^0_2\sim \tilde B$  there,  enforce unobservably  small values
for the $\tchi^0_1\tchi^0_2$ and $\tchi^0_2\tchi^0_2$
production cross sections.

\vspace{0.5cm}
We next turn to the comparison of the numerical results
from  the exact 1-loop computations, with those from the
asymptotic expressions in (\ref{leading-F},\ref{leading-LR}, \ref{leading-LI}).
First, we observed that in all benchmark models
considered above, the chirality violating
amplitude $F_{++++}$, and the chirality conserving
$F_{+-+-}$ and $F_{+--+}$, are the dominant  ones at energies above 1TeV.
In all cases, $F_{++++}$ is strongly decreasing with energy
in the several TeV range; while  $F_{+-+-}$ and $F_{+--+}$ become
increasingly dominant (with increasing energy) approaching
their  asymptotic expressions from (\ref{leading-F}), up to a constant.
Since (\ref{leading-F}) is generated by chargino-W loops,
the accuracy of the asymptotic expressions at high energies
 is particularly striking for cases where the produced neutralino
 is mainly  a Wino. A best example of these  is
illustrated in Fig.\ref{AD-fg5-1-amp-fig} for $\tchi_2\tchi_2$
production in the $AD(fg5)_1$ model.
There exist cases though, particularly in  the production of
 Bino-like neutralinos,  where the
 constant departure between exact and asymptotic expressions
for the $F_{+-+-}$ and $F_{+--+}$ amplitudes, and the subleading
  $F_{++++}$ amplitude, are still  important
  at the  few TeV energy region.\par

In practice it means that an amplitude analysis of
experimental data at high energy could immediately
determine the nature of the produced neutralinos.
If a precise amplitude analysis is not possible,
at least a fit of the unpolarized cross section
of the type $a + b \ln(s/m^2_W)+ c \ln^2(s/m^2_W)$
should produce a measurement
of the logarithmic slope $c$, which could then be compared to
the model predictions for the square of the quantities
$C_R$ or $C_I$ in (\ref{CRCI}), thereby  providing  clean tests
of the neutralino properties.

\section{Conclusions}

In this paper  we have presented our work on
the  properties of the process
$\gamma \gamma \to \tchi^0_i \tchi^0_j$, that are measurable
 in a $\gamma \gamma $ linear collider $LC_{\gamma \gamma}$.
It complements the study of the reverse process in \cite{nnDM3},
which was adjusted to the environment of the Dark Matter searches.
In both  cases, the complete set of  Feynman diagrams  has been calculated.

As in  \cite{nnDM3}, we have found  that these
processes are very sensitive to the actual values of the
various MSSM parameters, and
details of the neutralino mixings in particular.
The basic size of the cross sections can vary by several orders of
magnitude, depending on the contents of the produced neutralinos,
the highest values being obtained when
neutralinos are Wino-like. Depending on the SUSY model, these
properties reflect in a variable way in the
$\tchi^0_1~\tchi^0_2$ and $\tchi^0_2~\tchi^0_2$ productions.
Further enhancements can appear locally owing to the occurrence of
threshold effects due to relatively light intermediate states
in box diagrams or of $A^0$, $H^0$ resonance effects.

In the low and medium energy range these processes depend
on most of the MSSM parameters (because of the variety of
particles appearing as intermediate states in the one loop
diagrams of Fig.1) but we have also shown that,
in the high energy limit, the amplitudes tend
to a simple logarithmic form with a coefficient that only depends on
the neutralino mixing matrices
$Z^N$ without any other SUSY parameter.

The angular distribution of the differential cross sections
is generally rather flat at low energy,
but  a simple forward-backward peaking is progressively generated
according with the asymptotic rule  mentioned above.

These processes have  a very rich supersymmetric
contents (in fact all SUSY particles, except gluinos, contribute)
so that measurements at variable energy should provide very stringent
tests of  MSSM and  the neutralino structure.

Concerning in particular the Linear Collider measurements,
we should remember  that the $LC_{\gamma \gamma}$ study
of  the $\gamma \gamma \to \tchi^0_i \tchi^0_j$-process, complements
the study of  the tree level LC observables for
 $e^-e^+  \to \tchi^0_i \tchi^0_j$. If MSSM is really valid, and
 neutralinos turn out to be accessible to future LC and $LC_{\gamma \gamma}$,
 the combined  analyses of such experiments can  increase
 our knowledge on the nature of neutralinos and of SUSY in general.

But the most important fact is that they can be combined with the
dark matter searches, thus allowing exchange of knowledge between
accelerator particle physics and astrophysics.
To facilitate future work on LC or DM studies, the numerical
codes PLATONlc and PLATONdml have been released, calculating
the Linear Collider differential cross sections and the
Dark Matter rates respectively \cite{plato}. These  codes
 are applicable to any MSSM model with  real
parameters at the electroweak scale\footnote{By convention
we always select $M_2>0$.}.

\vspace{0.5cm}
\noindent
Acknowledgments:\\
It is a pleasure  to thank Abdelhak Djouadi and
Jean-Lo\"\i c Kneur for very helpful discussions.


\clearpage

\begin{figure}[p]
\vspace*{-3cm}
\[
\hspace{-1.cm}\epsfig{file=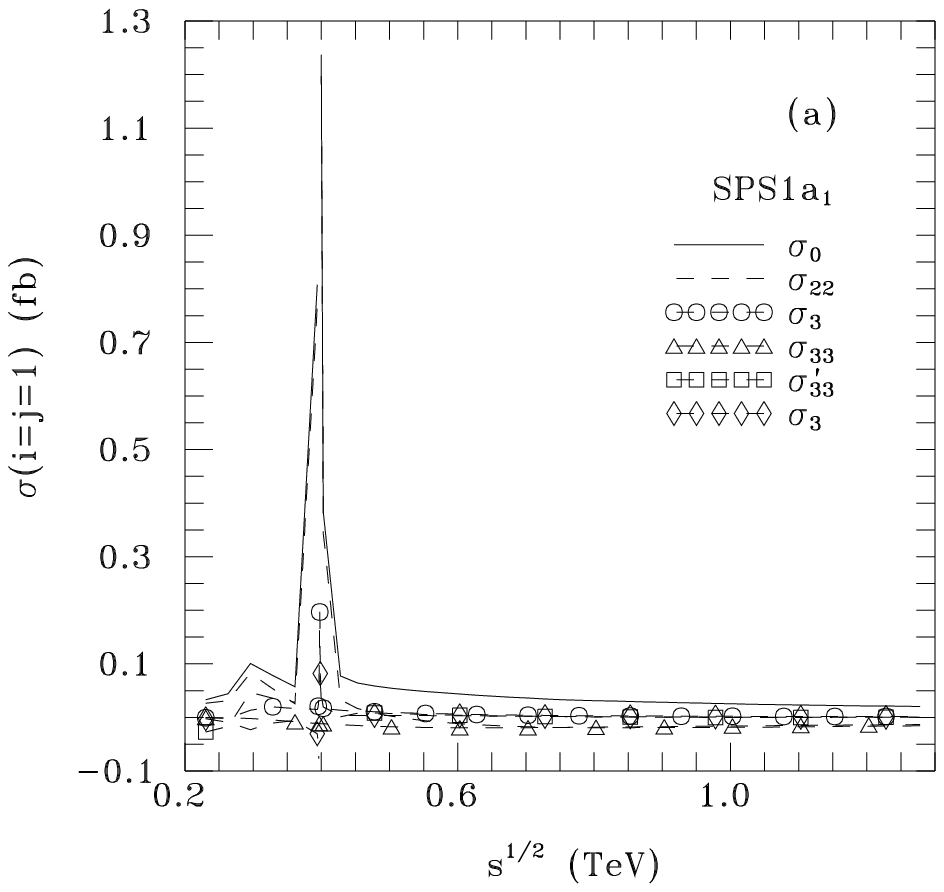,height=6.cm}
\hspace{1.cm}\epsfig{file=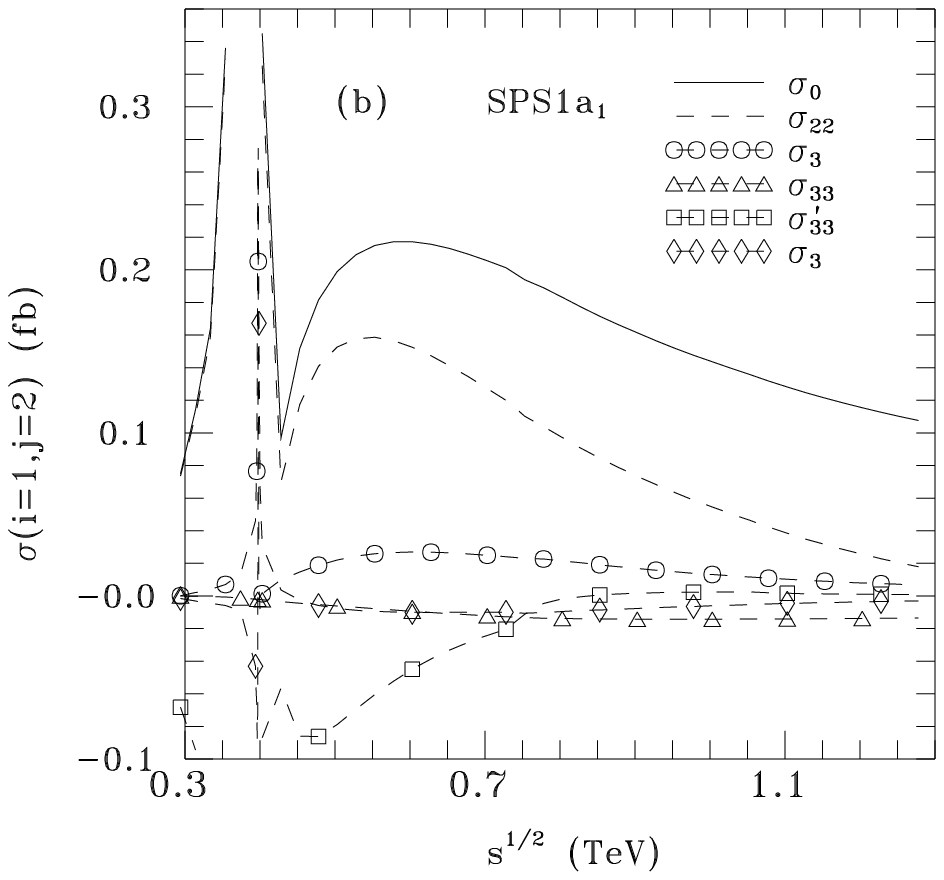,height=6.cm}
\]
\[
\hspace{-0.5cm}\epsfig{file=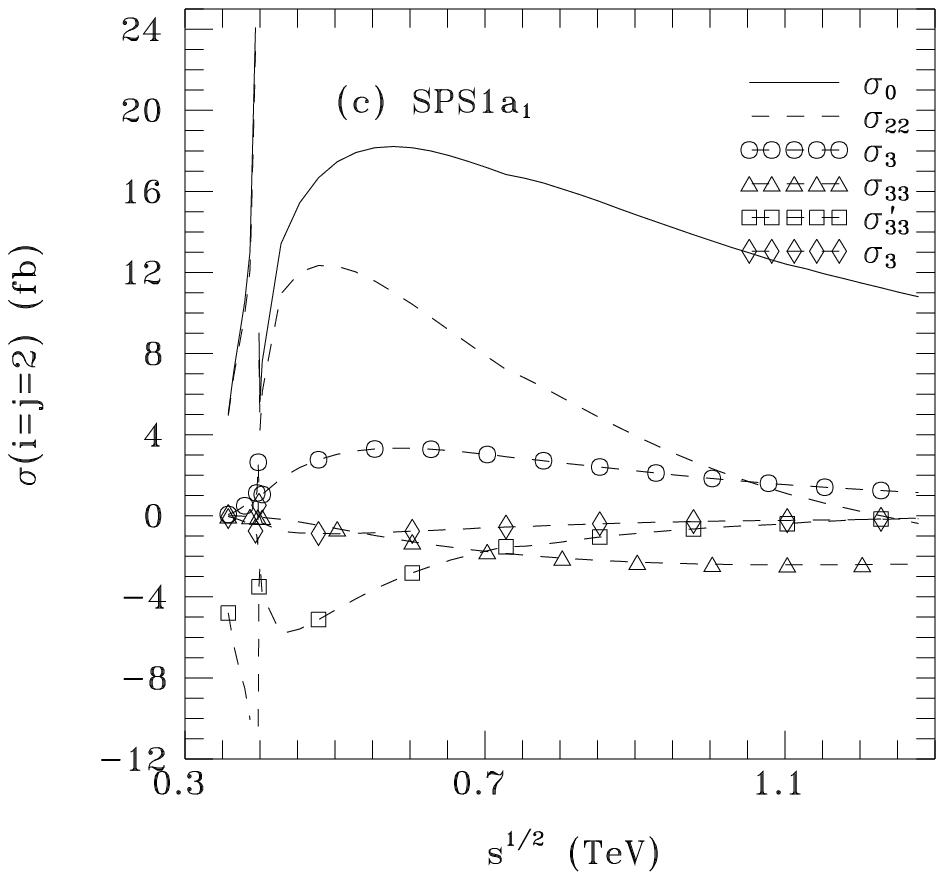,height=6.cm}
\]
\[
\hspace{-1.cm}\epsfig{file=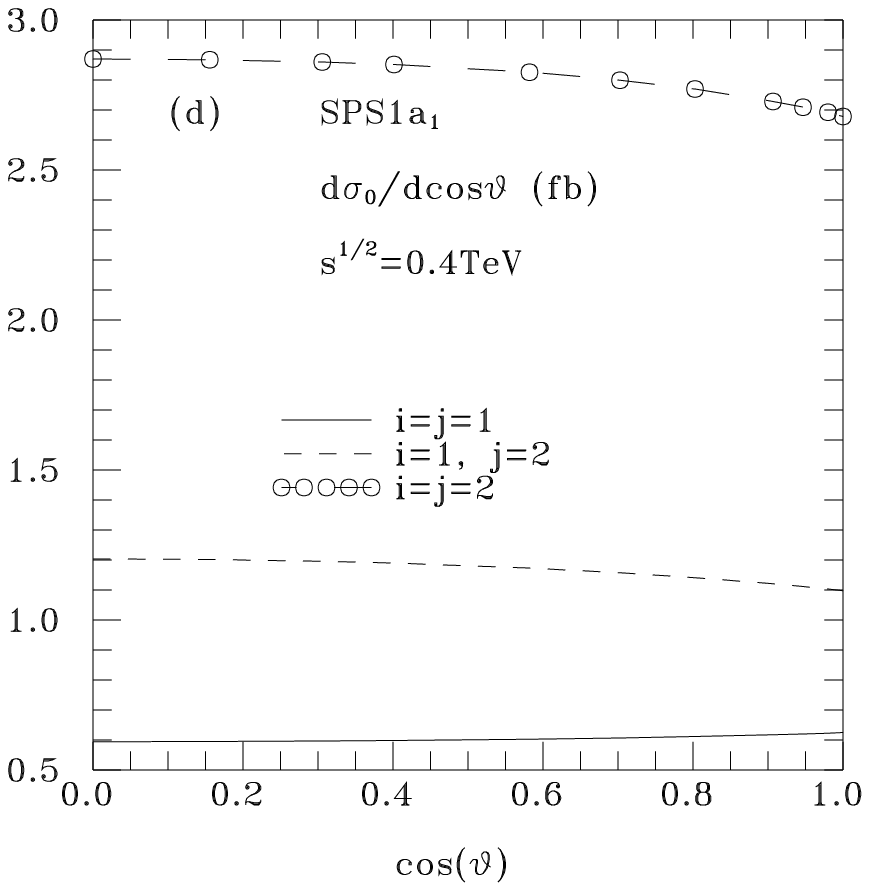,height=6.cm}
\hspace{1.cm}\epsfig{file=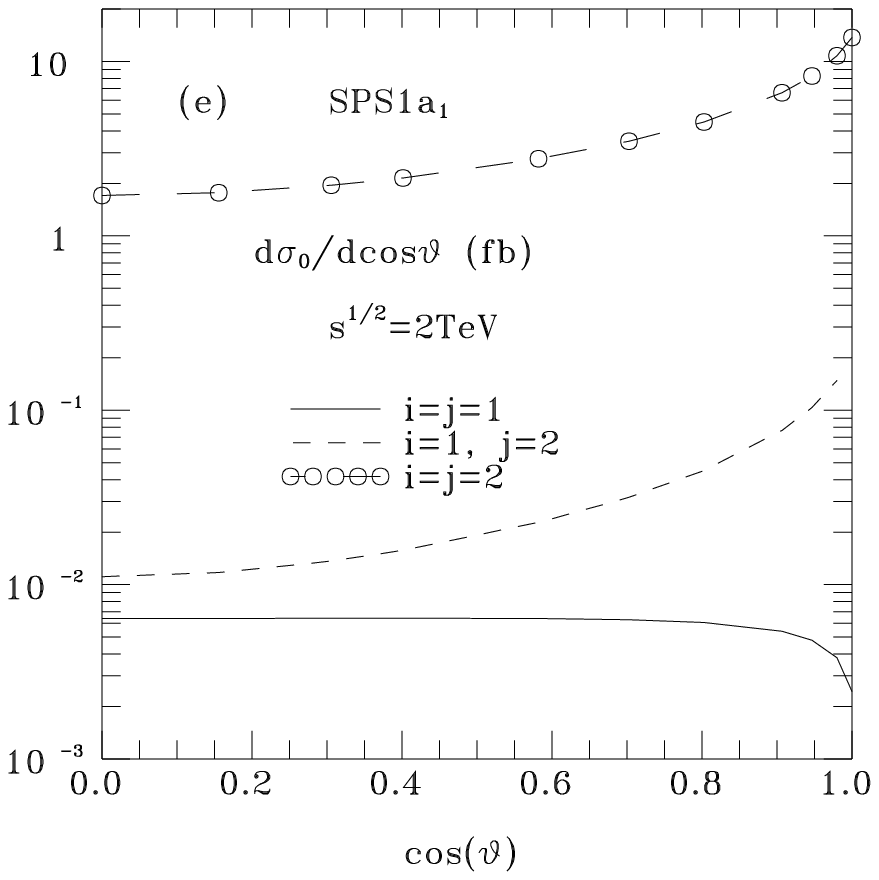,height=6.cm}
\]
\caption[1]{The integrated cross sections
for $\gamma\gamma \to \tchi^0_i \tchi^0_j$ in  the benchmark
model  $SPS1a_1$ \cite{Snowmass}
at a variable energy: i=1 j=1 (a), i=1 j=2 (b), i=2 j=2 (c). The
angular distributions for $\sigma_0$ in the same model are given
at 0.4 TeV (d) and 2 TeV (e). In all cases $\alpha=1/137$ is used.}
\label{SPS1a-1-fig}
\end{figure}

\clearpage

\begin{figure}[p]
\vspace*{-3cm}
\[
\hspace{-1.cm}\epsfig{file=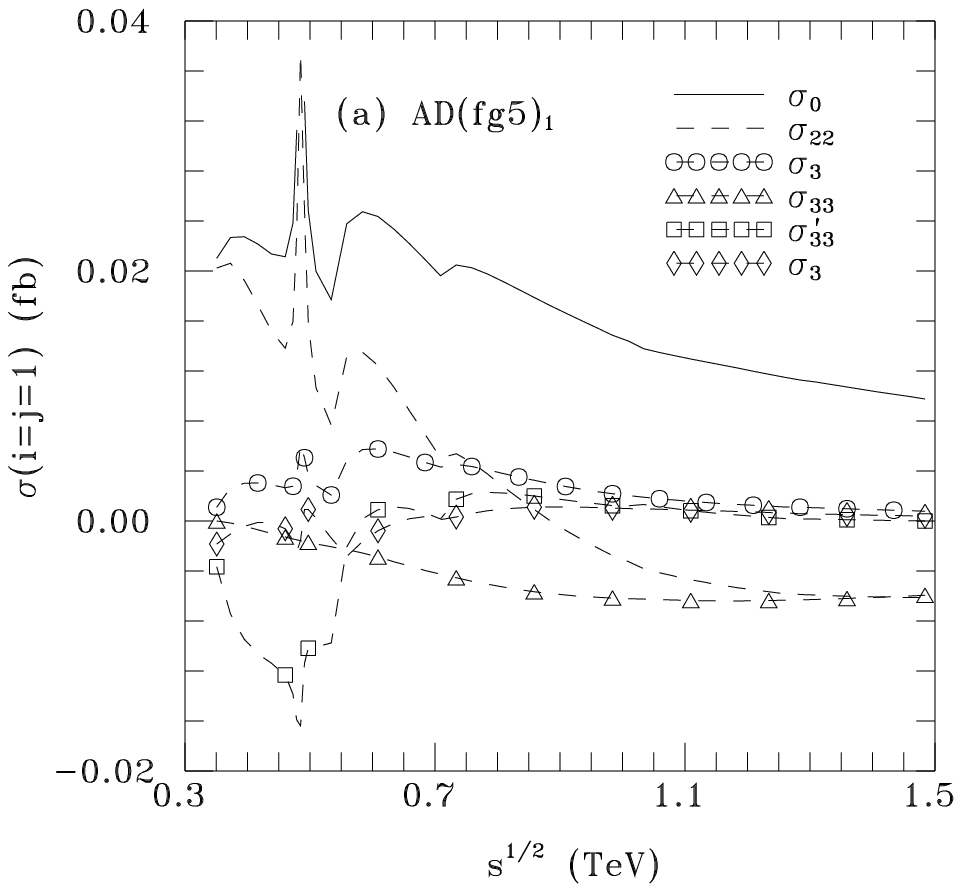,height=6.cm}
\hspace{1.cm}\epsfig{file=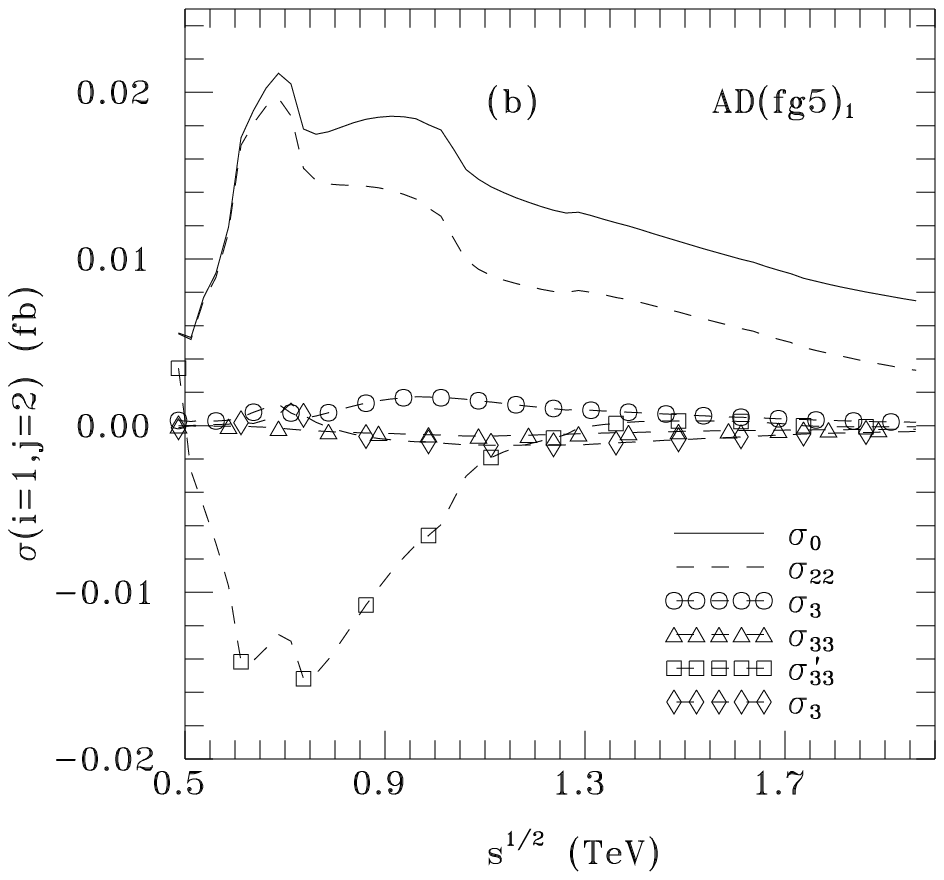,height=6.cm}
\]
\[
\hspace{-0.5cm}\epsfig{file=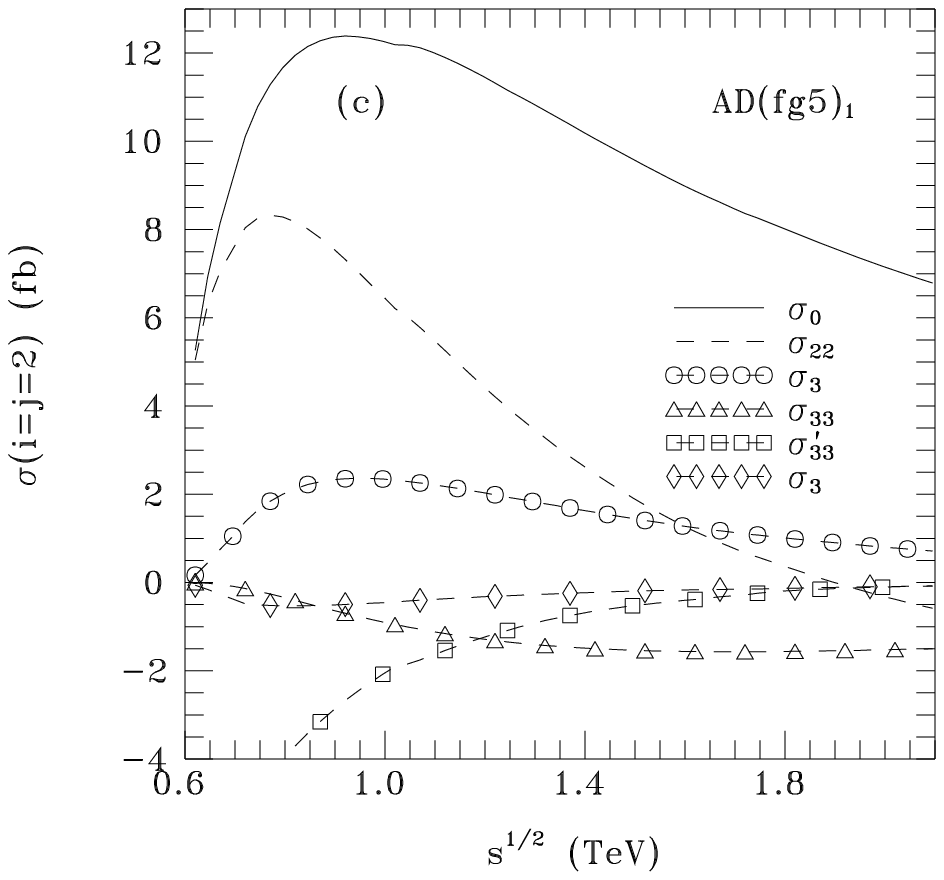,height=6.cm}
\]
\[
\hspace{-1.cm}\epsfig{file=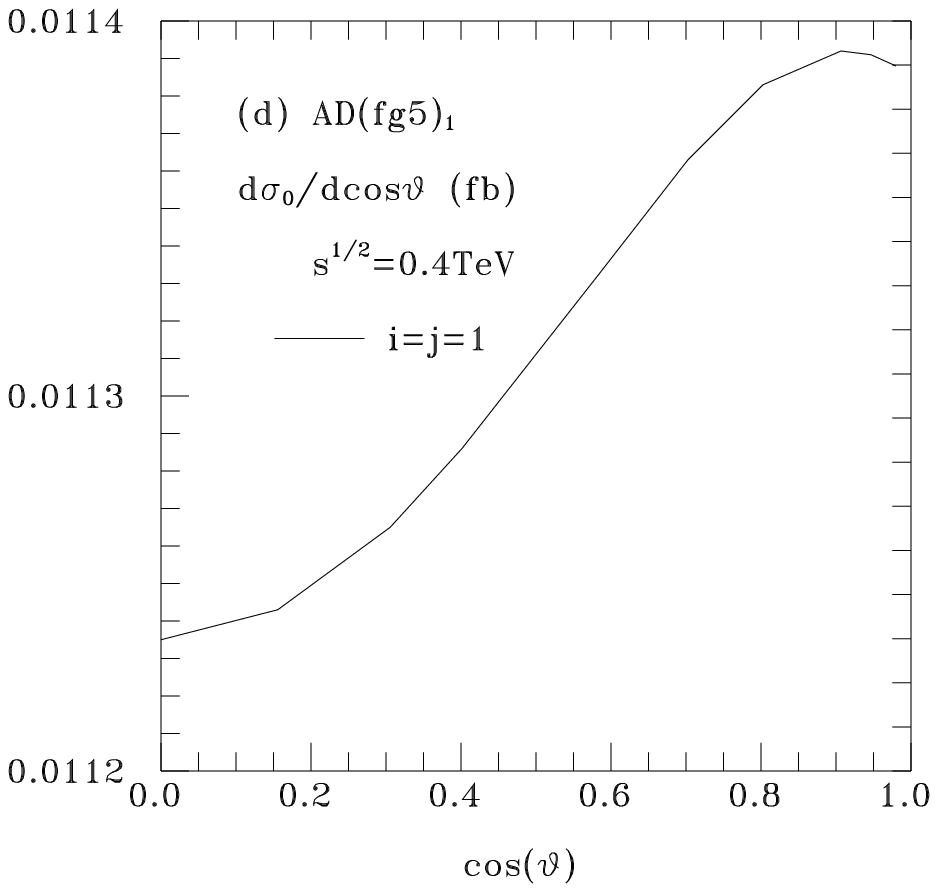,height=6.cm}
\hspace{1.cm}\epsfig{file=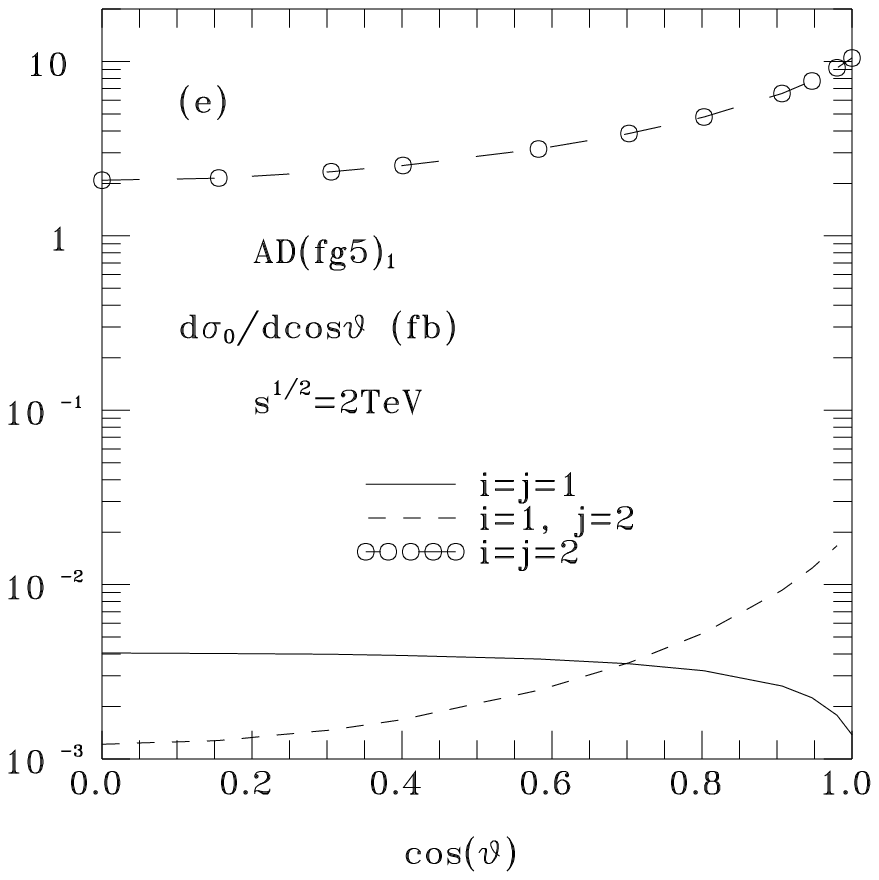,height=6.cm}
\]
\caption[1]{Same caption as in Fig.\ref{SPS1a-1-fig}
for  the benchmark model $AD(fg5)_1$ \cite{Arnowitt}.}
\label{AD-fg5-1-fig}
\end{figure}

\clearpage

\begin{figure}[p]
\vspace*{-3cm}
\[
\hspace{-1.cm}\epsfig{file=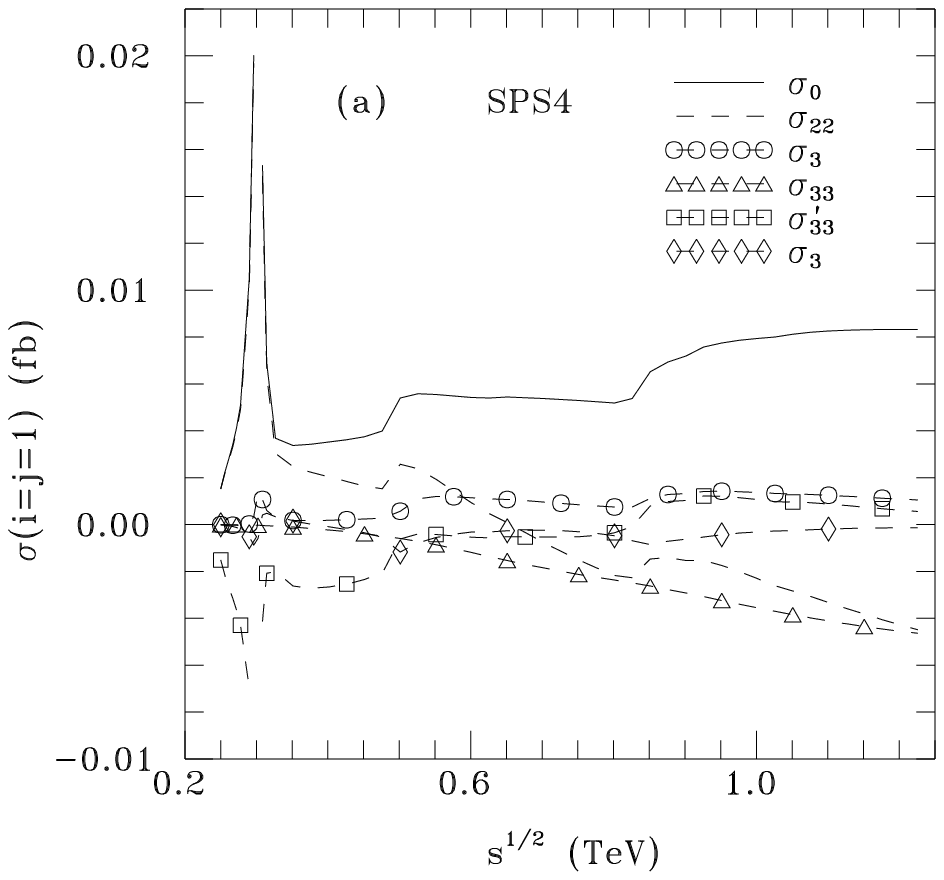,height=6.cm}
\hspace{1.cm}\epsfig{file=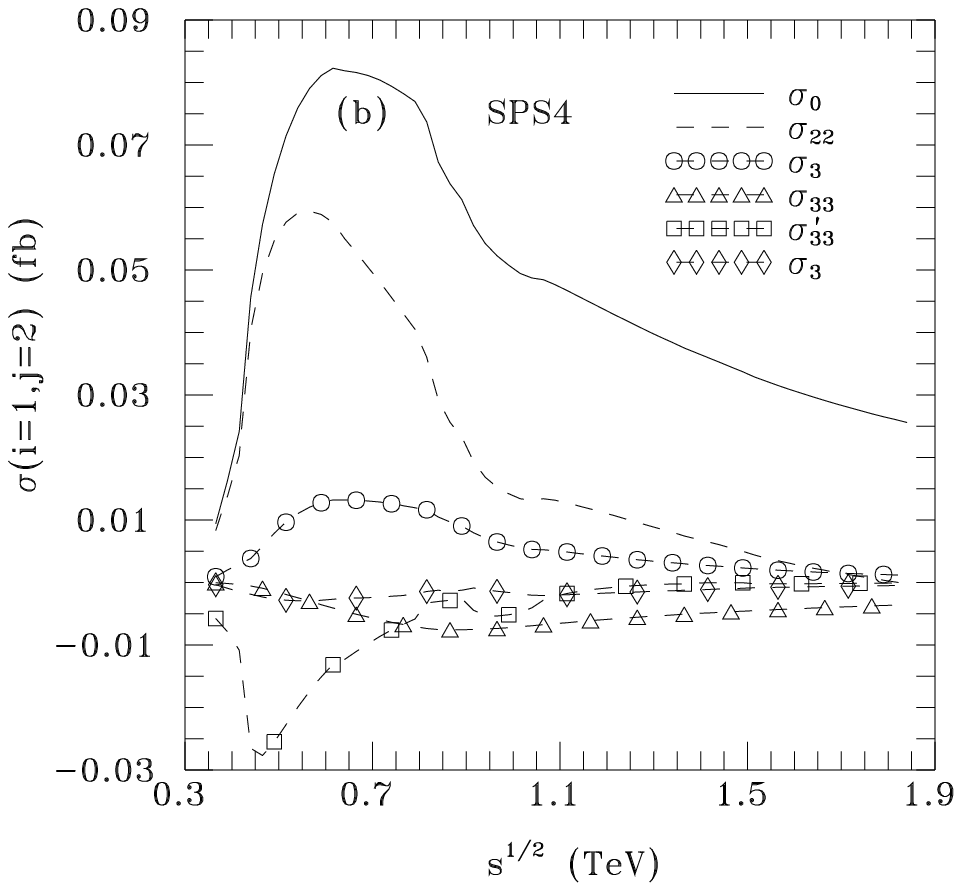,height=6.cm}
\]
\[
\hspace{-0.5cm}\epsfig{file=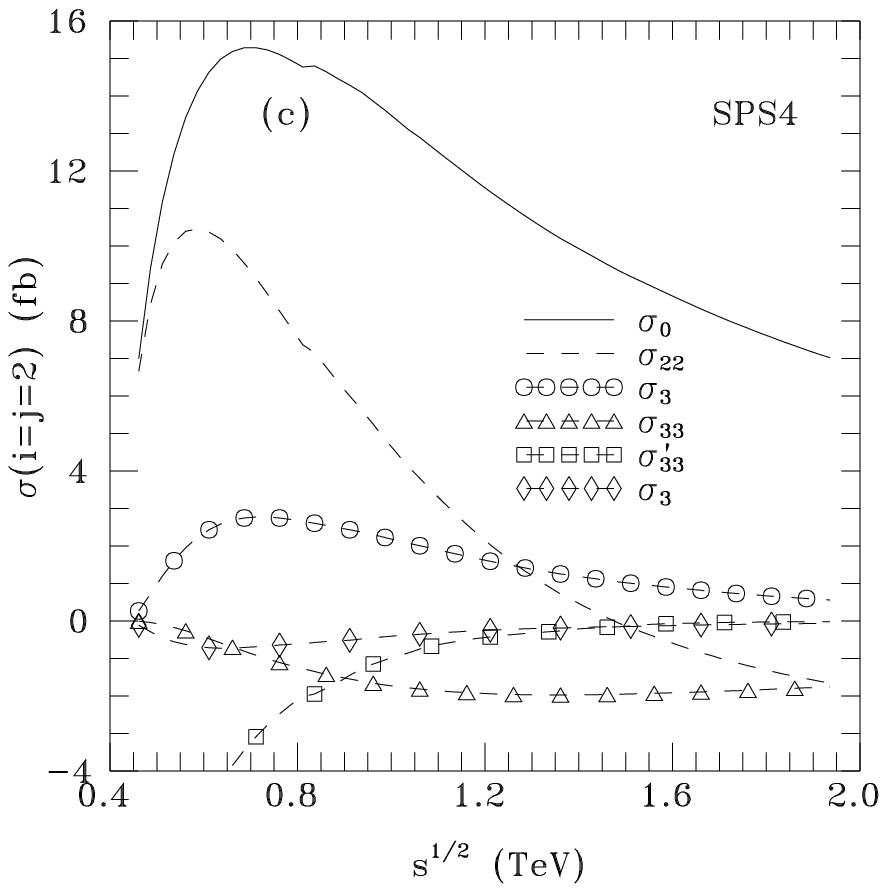,height=6.cm}
\]
\[
\hspace{-1.cm}\epsfig{file=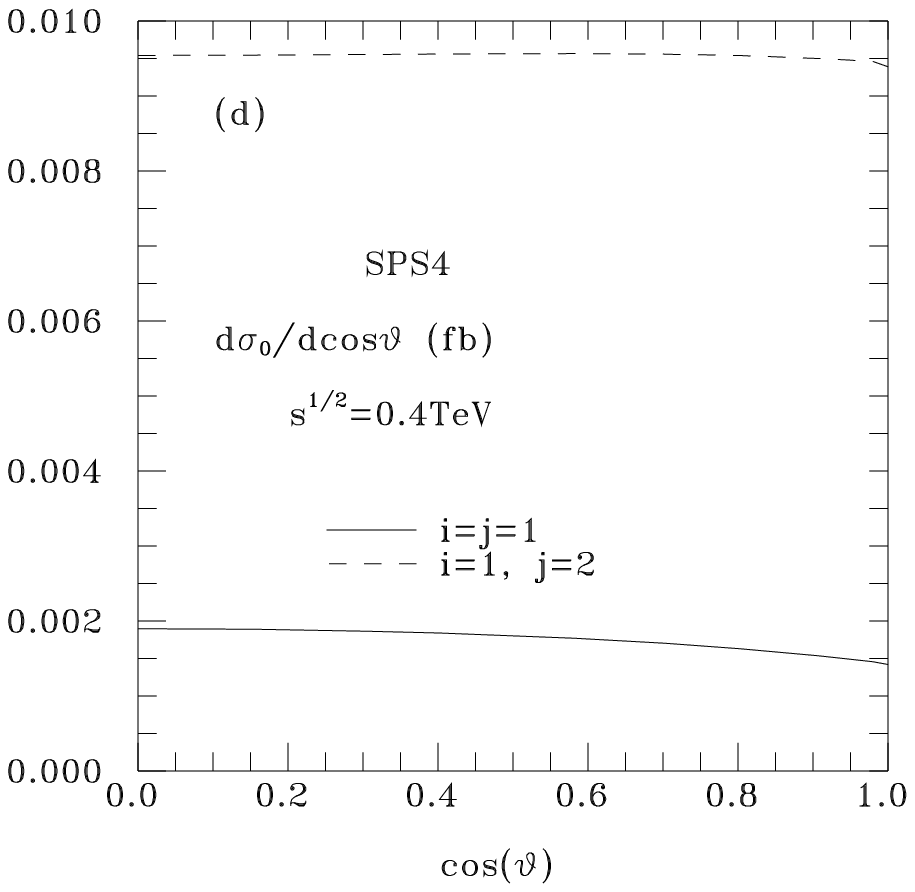,height=6.cm}
\hspace{1.cm}\epsfig{file=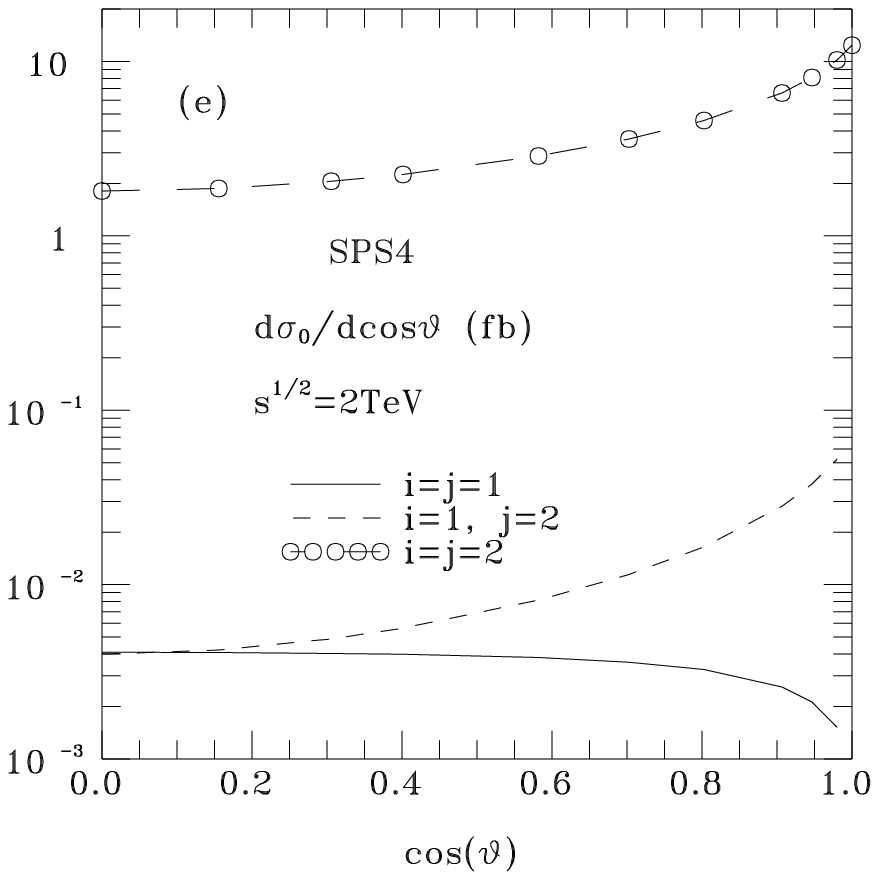,height=6.cm}
\]
\caption[1]{Same caption as in Fig.\ref{SPS1a-1-fig}
for  the benchmark model $SPS4$ \cite{Snowmass}. }
\label{SPS4-fig}
\end{figure}

\clearpage

\begin{figure}[p]
\vspace*{-3cm}
\[
\hspace{-1.cm}\epsfig{file=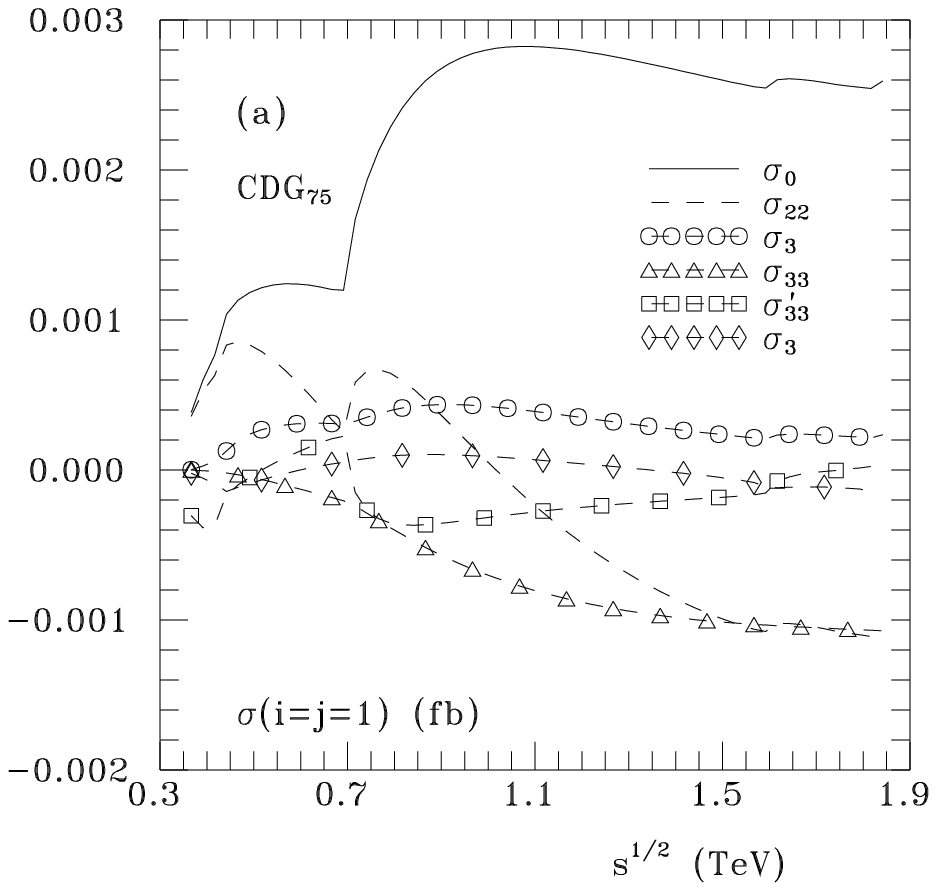,height=6.cm}
\hspace{1.cm}\epsfig{file=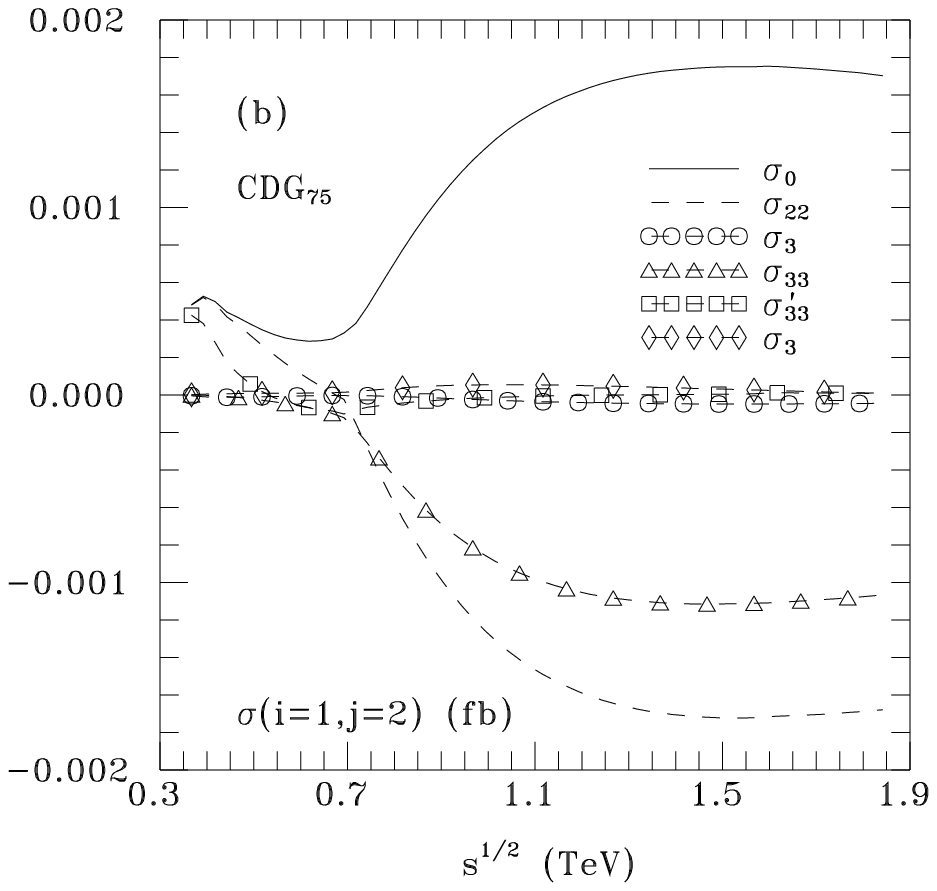,height=6.cm}
\]
\[
\hspace{-0.5cm}\epsfig{file=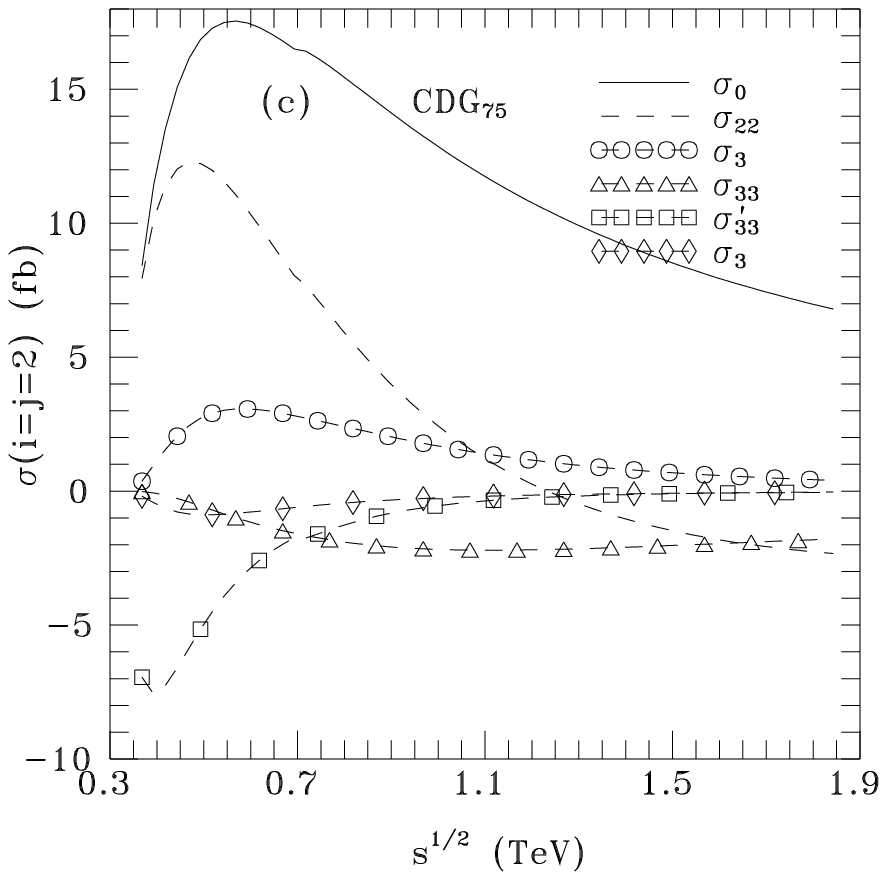,height=6.cm}
\]
\[
\hspace{-1.cm}\epsfig{file=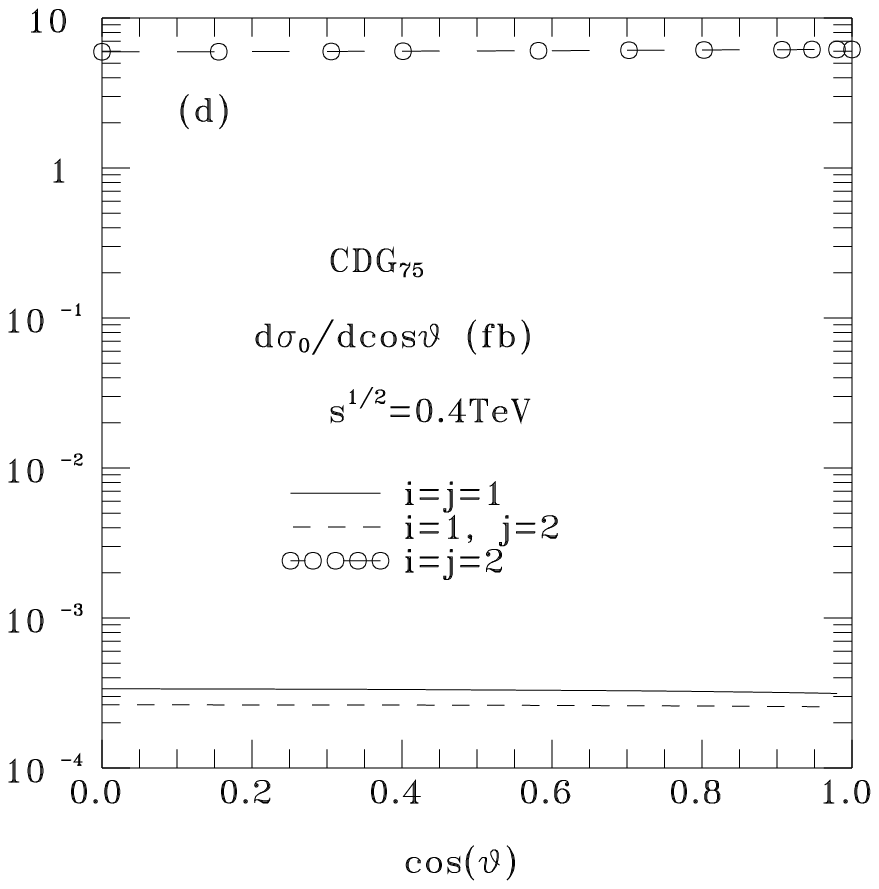,height=6.cm}
\hspace{1.cm}\epsfig{file=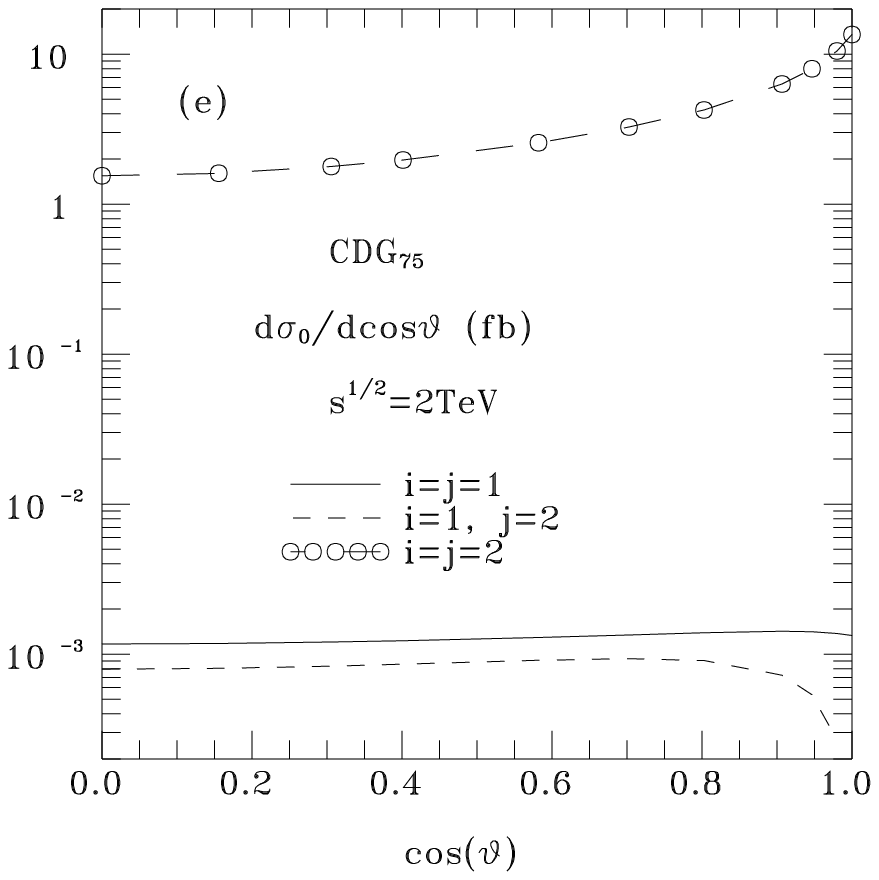,height=6.cm}
\]
\caption[1]{Same caption as in Fig.\ref{SPS1a-1-fig}
for  the benchmark model  $CDG_{75}$ \cite{CDG}. }
\label{CDG-75-fig}
\end{figure}

\clearpage

\begin{figure}[p]
\vspace*{-3cm}
\[
\hspace{-1.cm}\epsfig{file=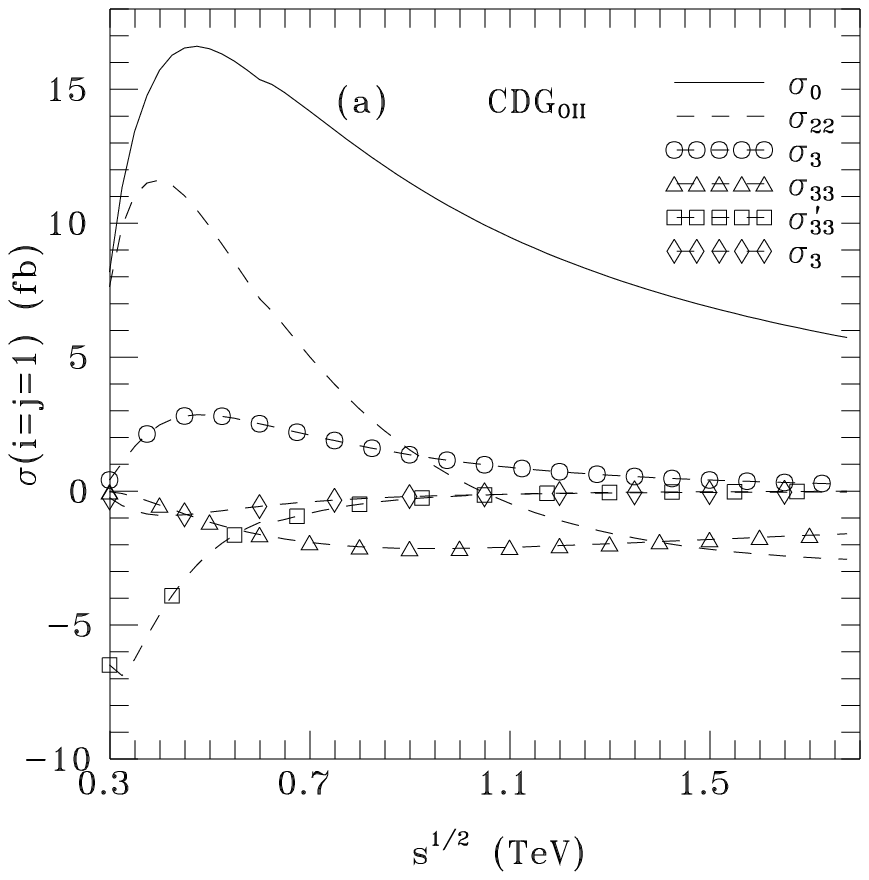,height=6.cm}
\hspace{1.cm}\epsfig{file=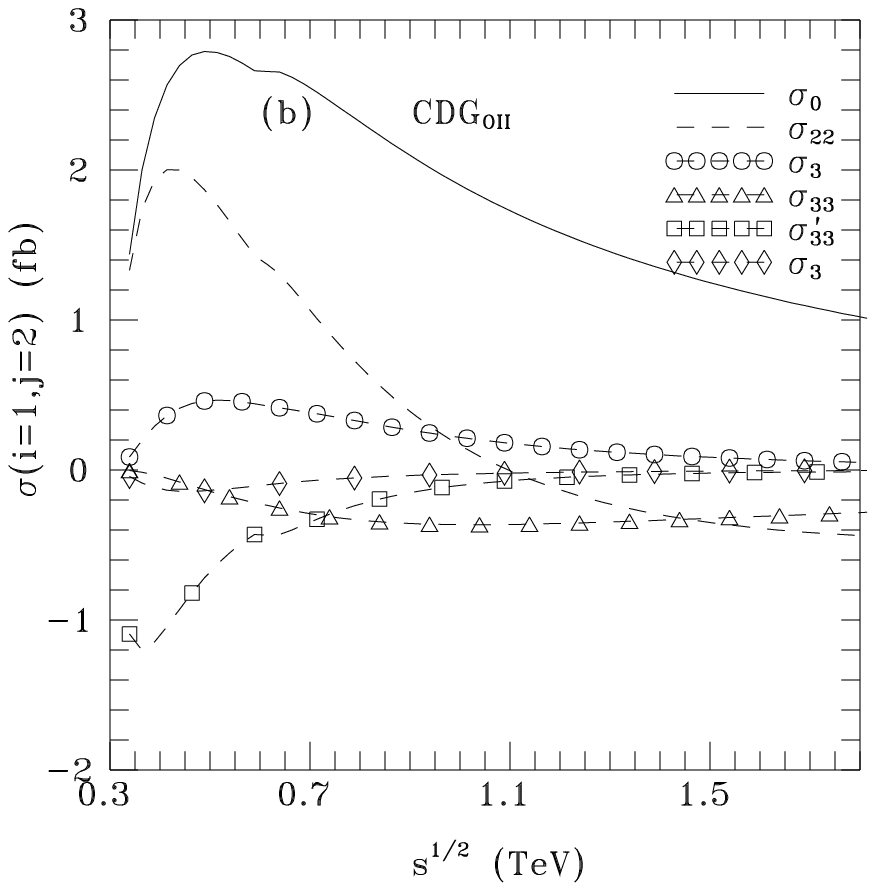,height=6.cm}
\]
\[
\hspace{-0.5cm}\epsfig{file=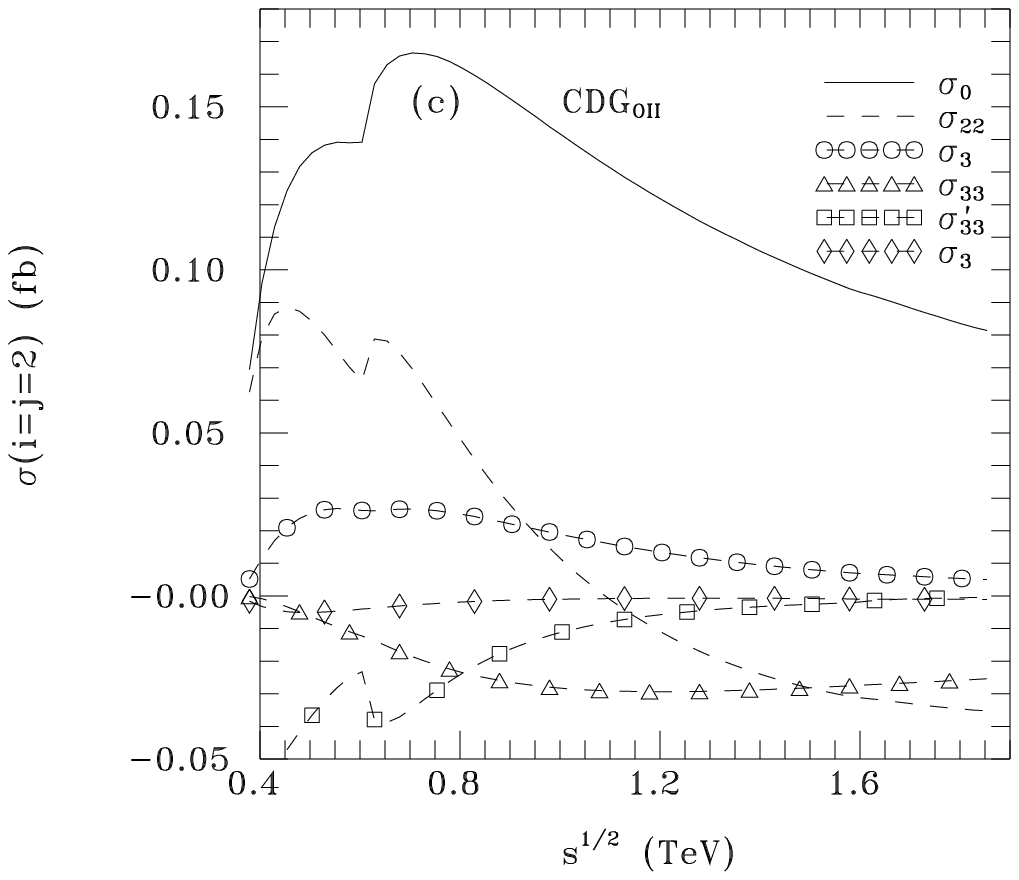,height=6.cm}
\]
\[
\hspace{-1.cm}\epsfig{file=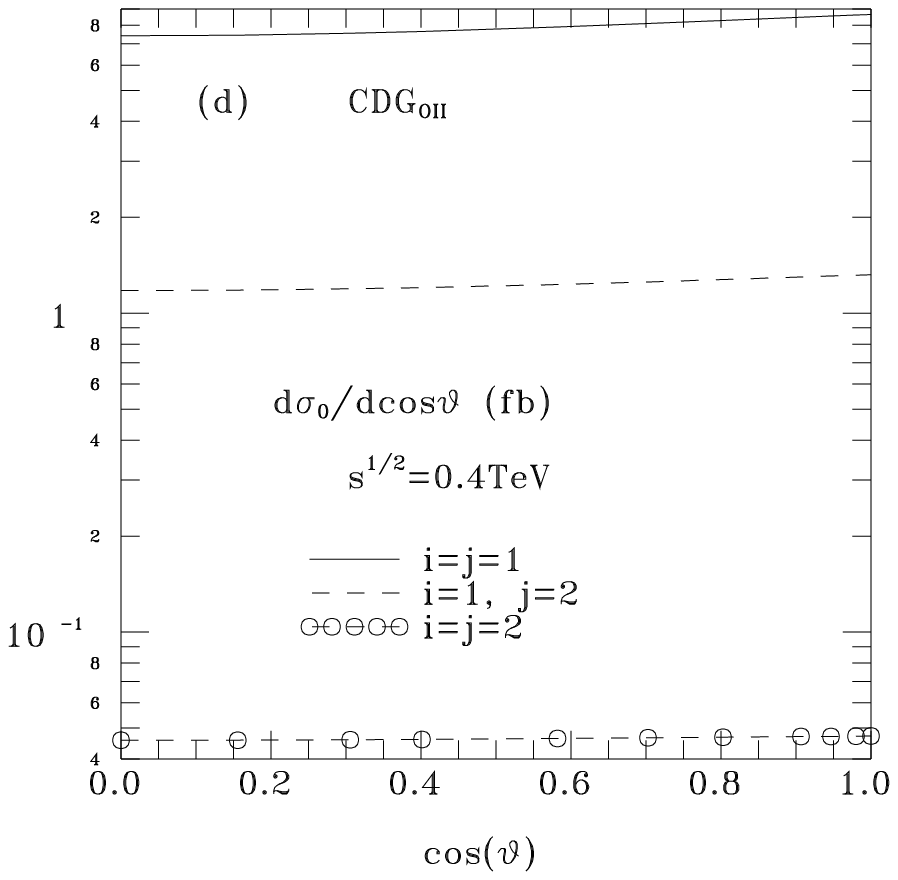,height=6.cm}
\hspace{1.cm}\epsfig{file=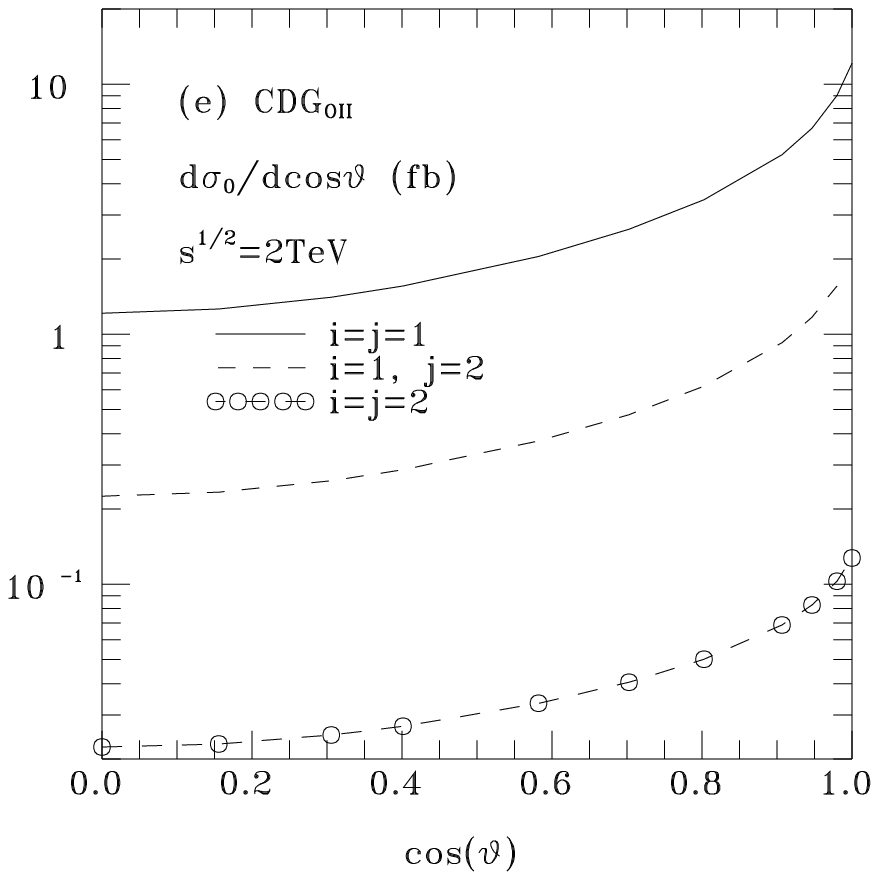,height=6.cm}
\]
\caption[1]{Same caption as in Fig.\ref{SPS1a-1-fig}
for  the benchmark model $CDG_{OII}$ \cite{CDG}. }
\label{CDG-OII-fig}
\end{figure}

\clearpage

\begin{figure}[p]
\vspace*{-3cm}
\[
\hspace{-1.cm}\epsfig{file=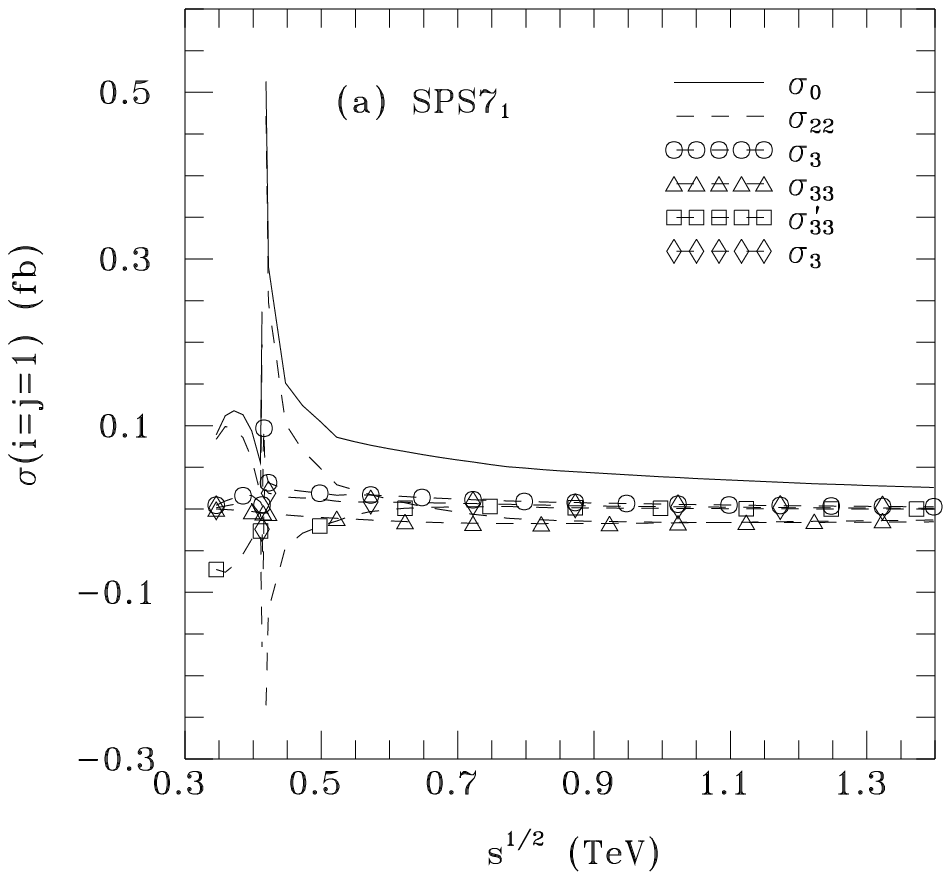,height=6.cm}
\hspace{1.cm}\epsfig{file=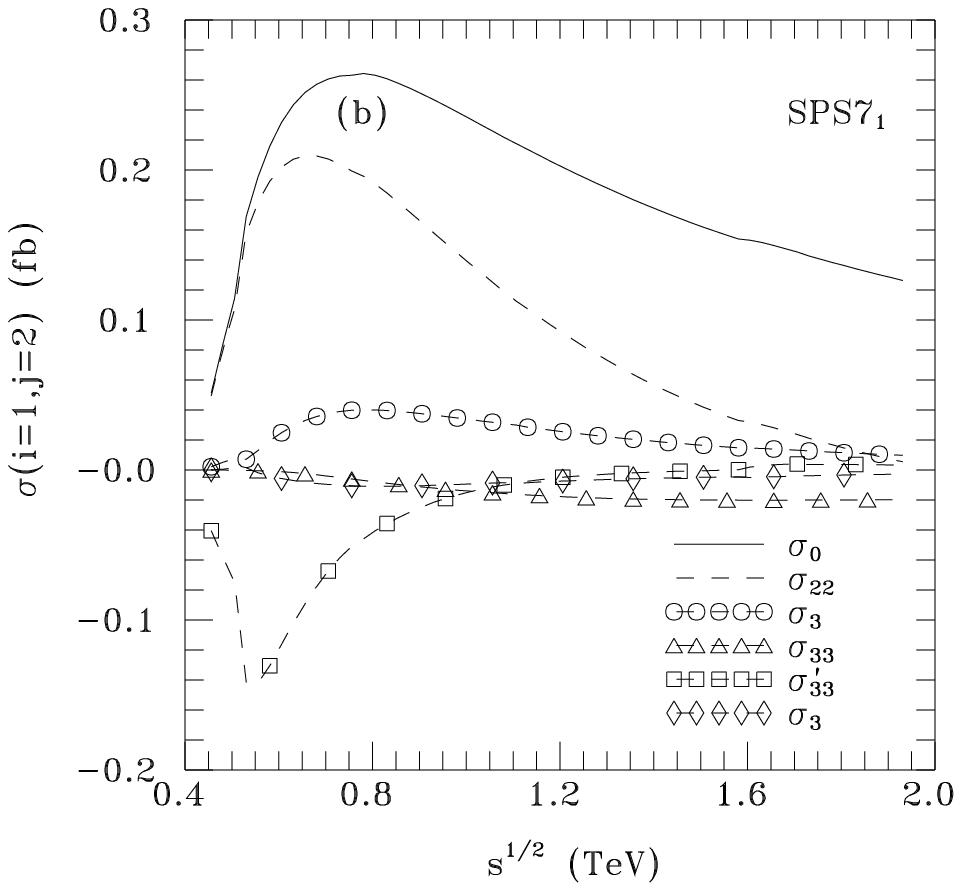,height=6.cm}
\]
\[
\hspace{-0.5cm}\epsfig{file=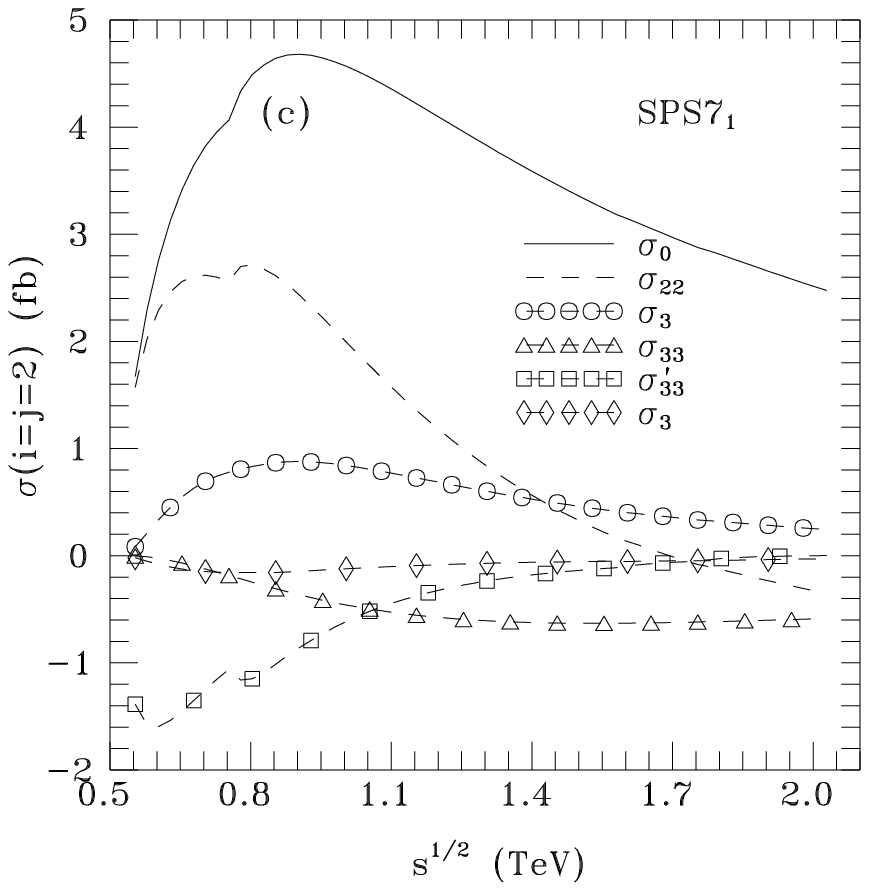,height=6.cm}
\]
\[
\hspace{-1.cm}\epsfig{file=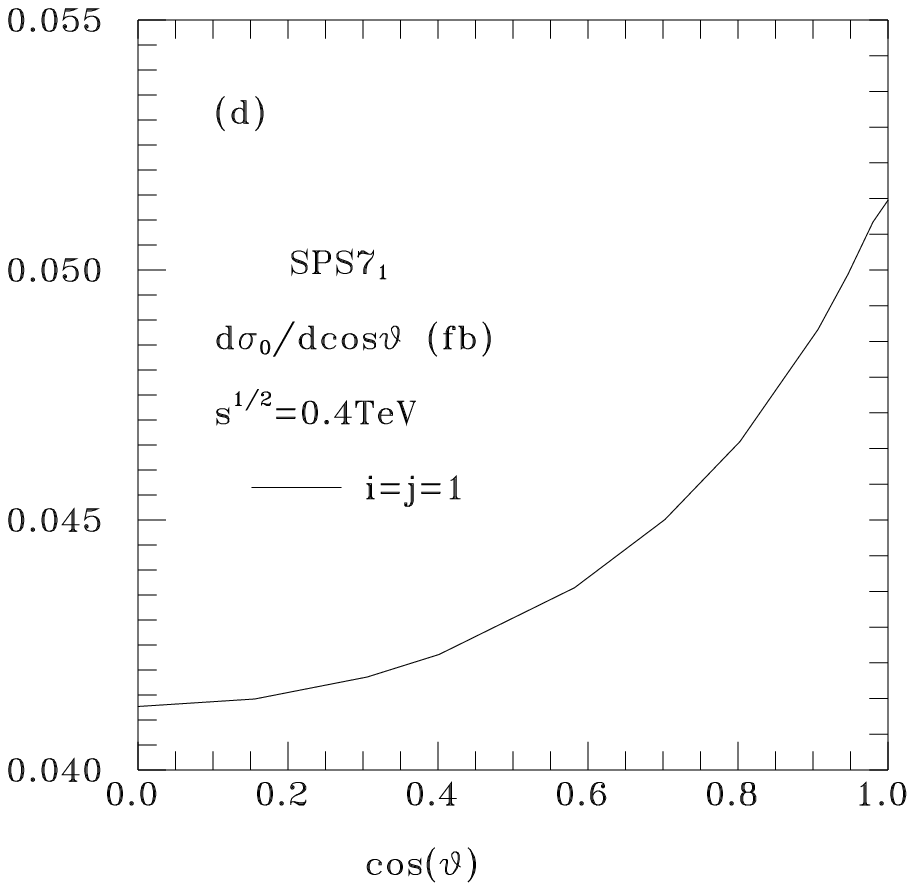,height=6.cm}
\hspace{1.cm}\epsfig{file=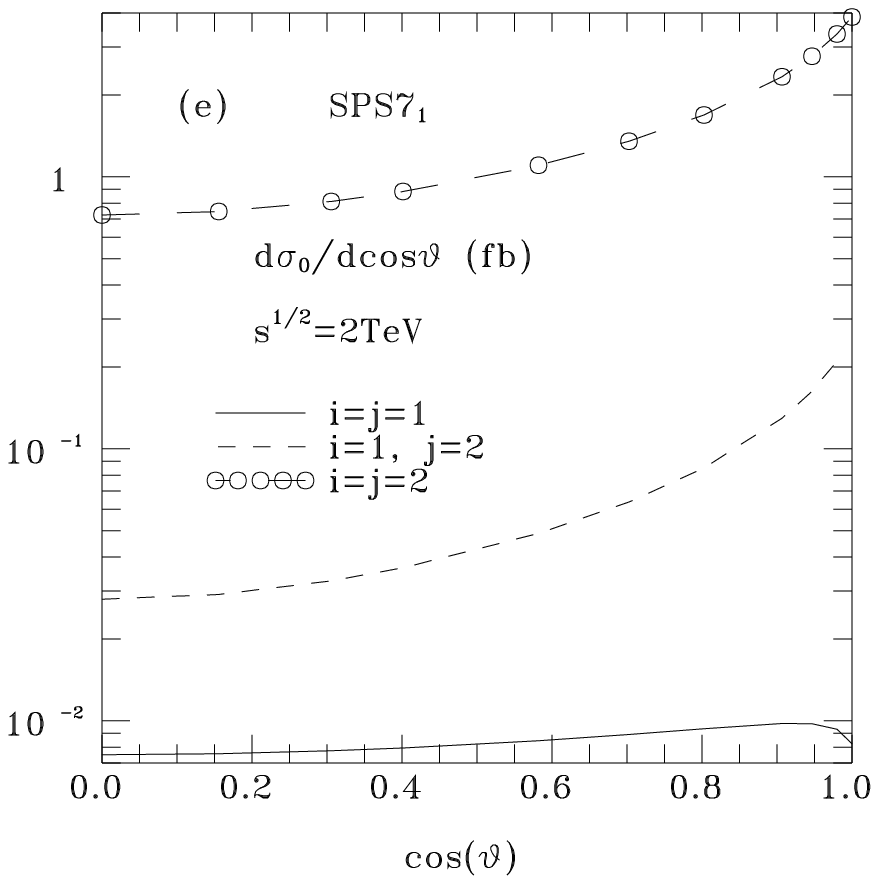,height=6.cm}
\]
\caption[1]{Same caption as in Fig.\ref{SPS1a-1-fig}
for  the benchmark model  $SPS7_1$ \cite{Snowmass}. }
\label{SPS7-1-fig}
\end{figure}

\clearpage

\begin{figure}[p]
\vspace*{-3cm}
\[
\hspace{-1.cm}\epsfig{file=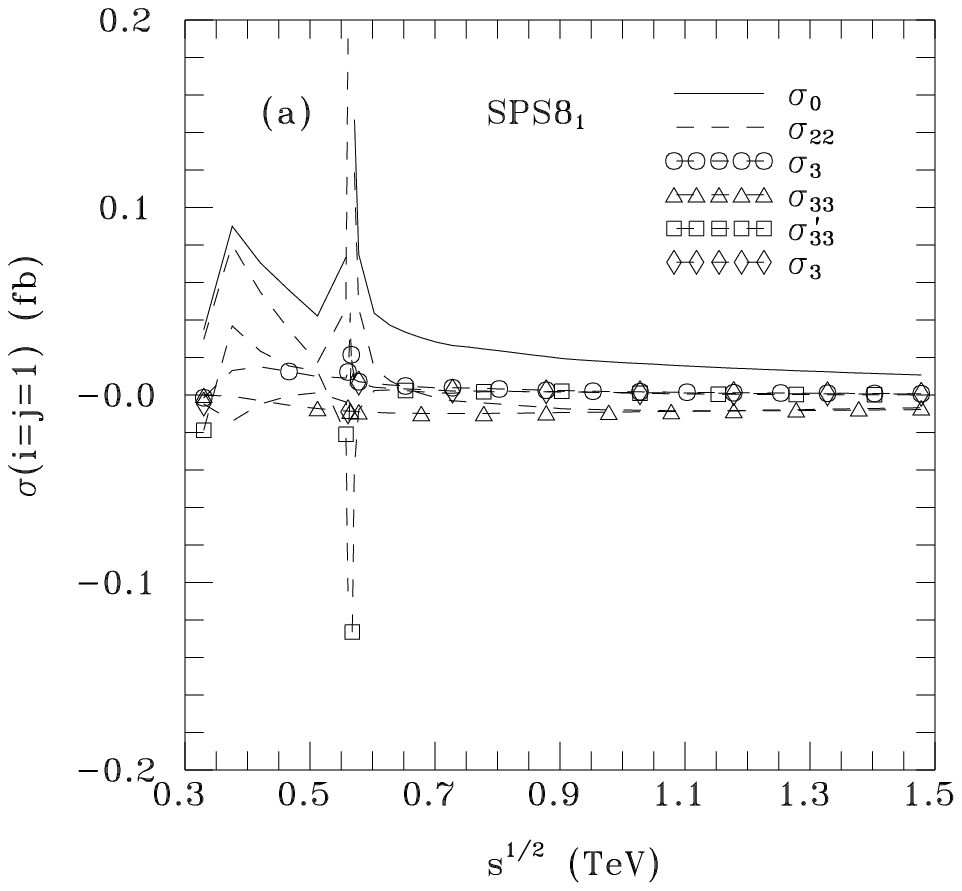,height=6.cm}
\hspace{1.cm}\epsfig{file=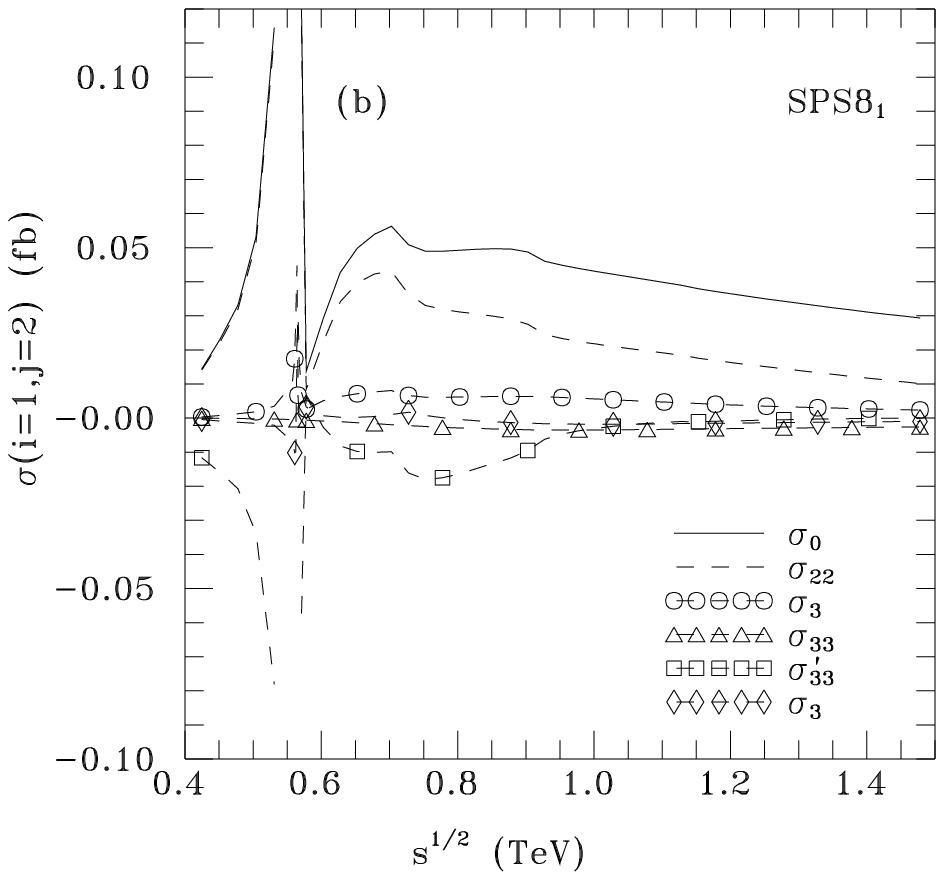,height=6.cm}
\]
\[
\hspace{-0.5cm}\epsfig{file=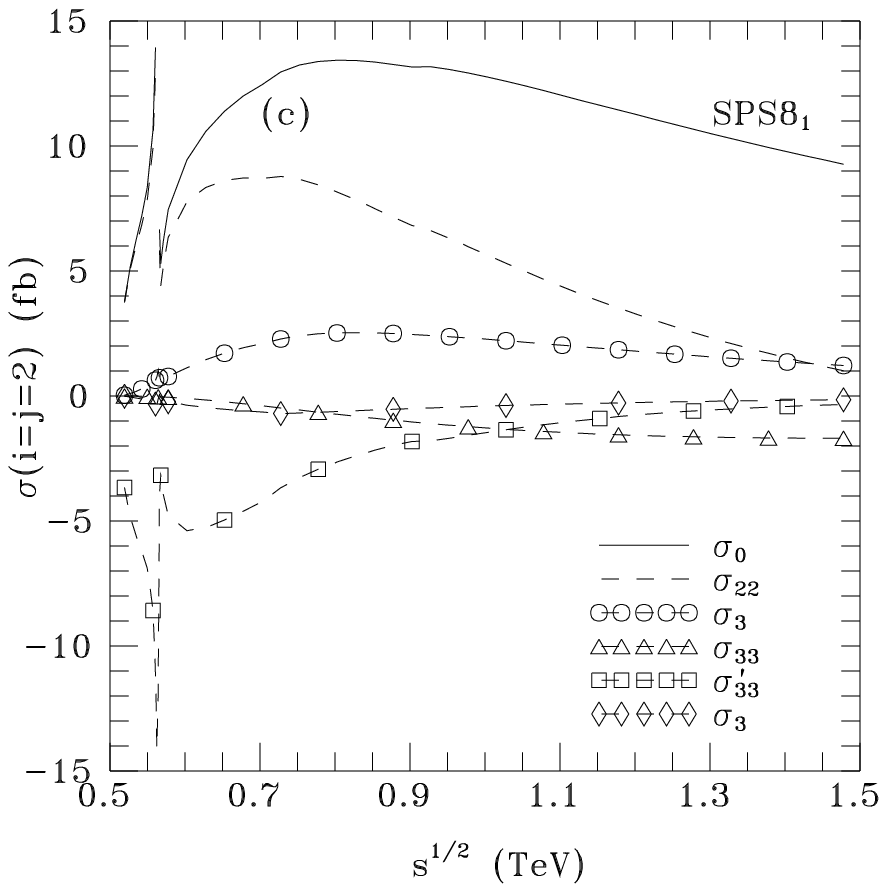,height=6.cm}
\]
\[
\hspace{-1.cm}\epsfig{file=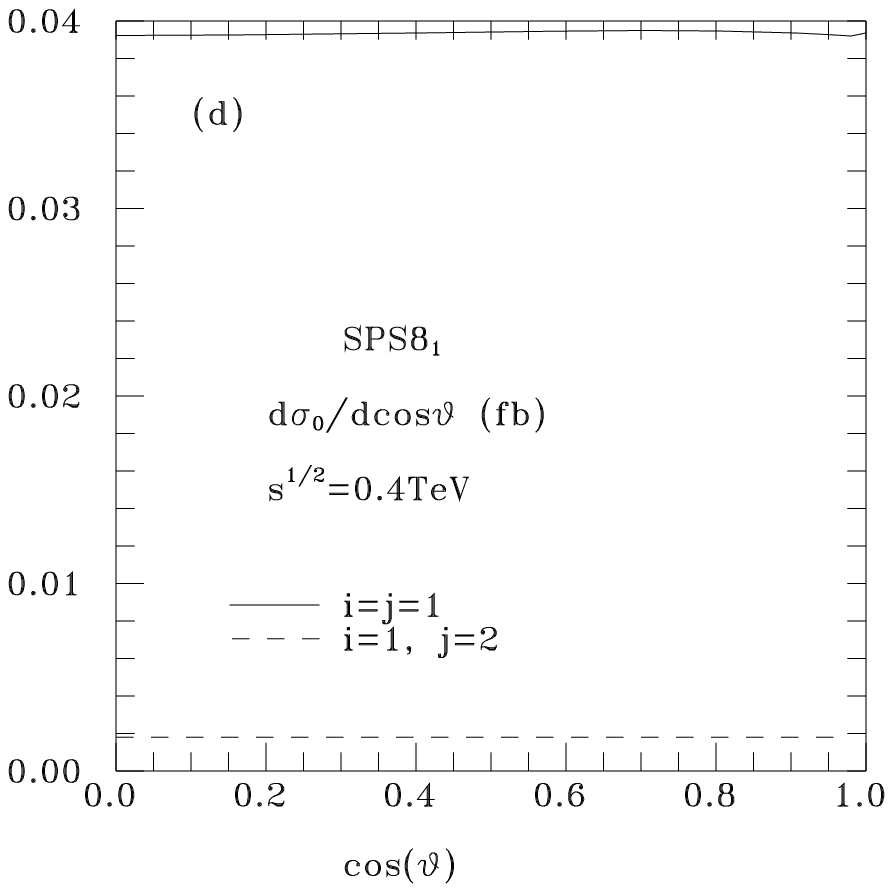,height=6.cm}
\hspace{1.cm}\epsfig{file=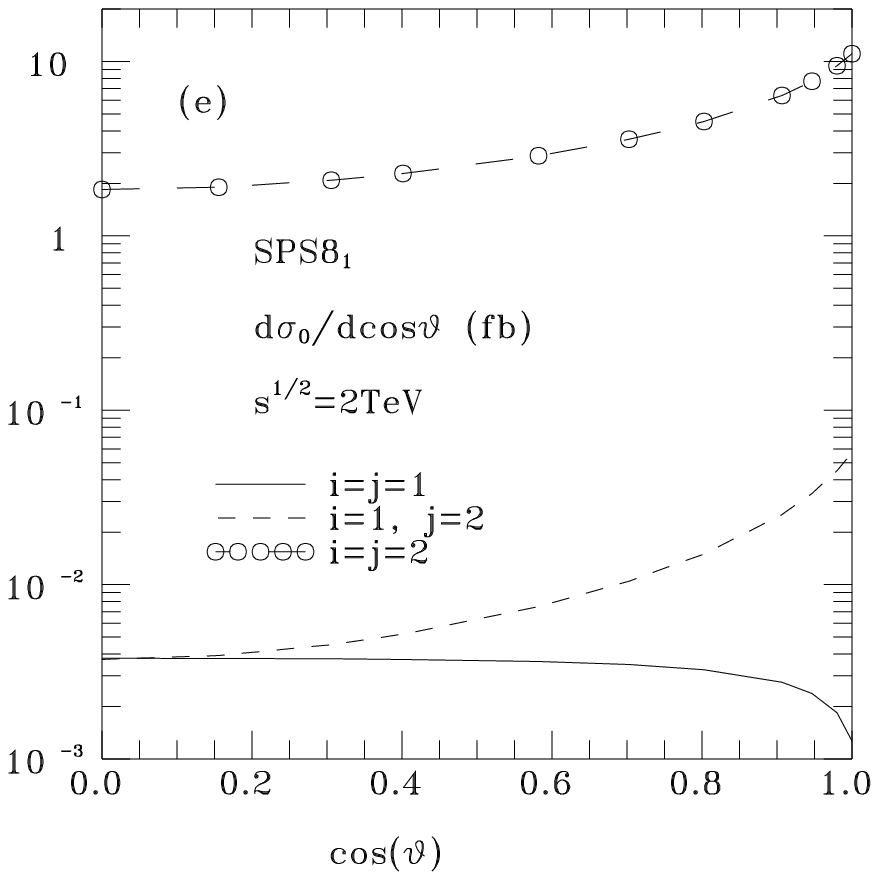,height=6.cm}
\]
\caption[1]{Same caption as in Fig.\ref{SPS1a-1-fig}
for  the benchmark model  $SPS8_1$ \cite{Snowmass}.}
\label{SPS8-1-fig}
\end{figure}

\begin{figure}[p]
\vspace*{-3cm}
\[
\hspace{-1.cm}\epsfig{file=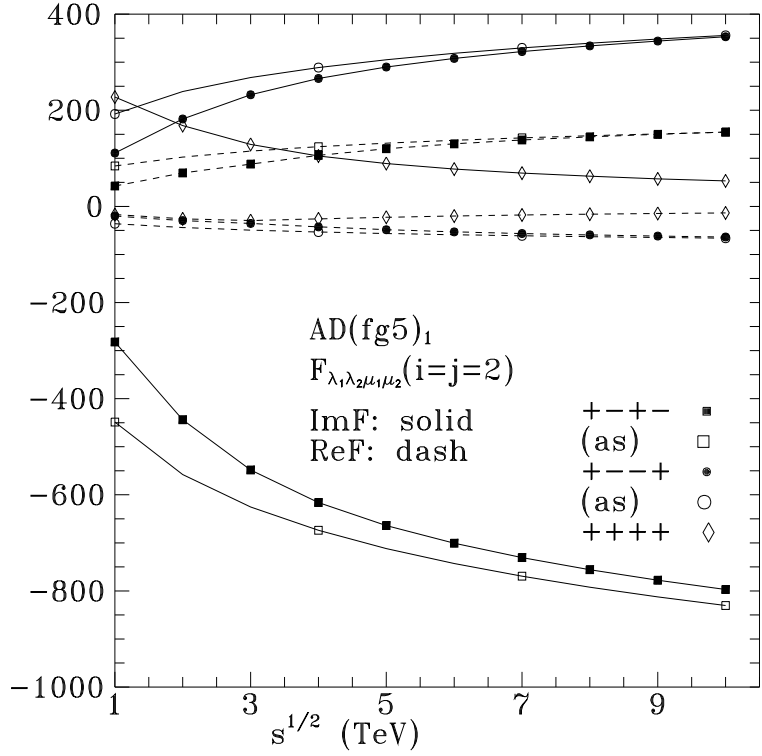,height=10.cm, width=10.cm}
\]
\caption[1]{The exact  and asymptotic (indicated by "as" ) 1-loop
expressions for the real and imaginary parts of the $F_{+-+-}$, $F_{+--+}$
and $F_{++++}$ helicity amplitudes (with a factor $\alpha^2$ removed)
at $\theta=\pi/4$, for the $i=j=2$ case in the
$AD(fg5)_1$-model \cite{Arnowitt}.}
\label{AD-fg5-1-amp-fig}
\end{figure}

\end{document}